\documentclass[journal]{IEEEtran}
\ifCLASSINFOpdf
\else
\fi
\usepackage[cmex10]{amsmath}
\newcommand{\eg}{\textit{e.g.}}
\newcommand{\ie}{\textit{i.e.}}
\newcommand{\cf}{\textit{cf.}}
\newcommand{\etal}{\textit{et~al.}}

\usepackage[T1]{fontenc}
\usepackage[utf8]{inputenc}

\usepackage{amsfonts,amssymb,amsthm,mathrsfs}
\usepackage{bm}
\newcommand{\mathbold}[1]{\bm{#1}}
\newcommand{\mbf}[1]{\mathbf{#1}}
\newcommand{\vect}[1]{\mbf{#1}}

\newcommand{\T}{\mathsf{T}}    %
\newcommand{\dd}{\,\mathrm{d}} %
\newcommand{\E}{\mathbb{E}}    %
\newcommand{\R}{\mathbb{R}}    %
\newcommand{\N}{\mathrm{N}}    %
\DeclareMathOperator{\tr}{tr}
\DeclareMathOperator{\diag}{diag}

\DeclareMathOperator{\argmax}{arg\,max}

\newcommand{\vmu}[0]{\mathbold{\mu}}

\newcommand{\vxi}[0]{\mathbold{\xi}}
\newcommand{\vtheta}[0]{\mathbold{\theta}}

\newcommand{\MPhi}[0]{\mathbold{\Phi}}
\newcommand{\MSigma}[0]{\mathbold{\Sigma}}
\newcommand{\MTheta}[0]{\mathbold{\Theta}}

\newcommand{\rc}{\mathrm{c}}

\newcommand{\ve}{\mbf{e}}
\newcommand{\vf}{\mbf{f}}
\newcommand{\vg}{\mbf{g}}
\newcommand{\vh}{\mbf{h}}

\newcommand{\vm}{\mbf{m}}

\newcommand{\vq}{\mbf{q}}
\newcommand{\vr}{\mbf{r}}
\newcommand{\vs}{\mbf{s}}

\newcommand{\vv}{\mbf{v}}

\newcommand{\vx}{\mbf{x}}
\newcommand{\vy}{\mbf{y}}

\newcommand{\MA}{\mbf{A}}
\newcommand{\MB}{\mbf{B}}
\newcommand{\MC}{\mbf{C}}
\newcommand{\MD}{\mbf{D}}
\newcommand{\MF}{\mbf{F}}
\newcommand{\MG}{\mbf{G}}
\newcommand{\MH}{\mbf{H}}

\newcommand{\MK}{\mbf{K}}
\newcommand{\ML}{\mbf{L}}

\newcommand{\MP}{\mbf{P}}
\newcommand{\MQ}{\mbf{Q}}
\newcommand{\MR}{\mbf{R}}
\newcommand{\MS}{\mbf{S}}

\usepackage{tikz,pgfplots}
\usetikzlibrary{plotmarks}

\newlength{\figurewidth}
\newlength{\figureheight}

\ifCLASSOPTIONcompsoc
  \usepackage[caption=false,font=normalsize,labelfont=sf,textfont=sf]{subfig}
\else
  \usepackage[caption=false,font=scriptsize]{subfig} %
\fi
\usepackage{color}

\usepackage[numbers,sort&compress]{natbib}

\begin{document}
\title{Sigma-Point Filtering and Smoothing Based Parameter Estimation in Nonlinear Dynamic Systems}

\author{Juho~Kokkala,
        Arno~Solin,
        and~Simo~S\"arkk\"a%
\thanks{The authors are with Aalto University, Espoo, Finland. Email: juho.kokkala@aalto.fi, arno.solin@aalto.fi, simo.sarkka@aalto.fi.}%
\thanks{This work was supported by grants from the Academy of Finland (266940, 273475) and by the Emil Aaltonen foundation. }
\thanks{The authors declare no conflict-of-interest.}
\thanks{Manuscript received March 31, 2015. Revised August 31, 2015.}} %

\markboth{Submitted to Journal of Advances in Information Fusion}%
{Kokkala \MakeLowercase{\textit{et al.}}: Sigma-Point Filtering...}
\maketitle

\begin{abstract} %
We consider approximate maximum likelihood parameter estimation in nonlinear state-space models. We discuss both direct optimization of the likelihood and expectation--maximization (EM). For EM, we also give closed-form expressions for the maximization step in a class of models that are linear in parameters and have additive noise. To obtain approximations to the filtering and smoothing distributions needed in the likelihood-maximization methods, we focus on using Gaussian filtering and smoothing algorithms that employ sigma-points to approximate the required integrals. We discuss different sigma-point schemes based on the third, fifth, seventh, and ninth order unscented transforms and the Gauss--Hermite quadrature rule. We compare the performance of the methods in two simulated experiments: a univariate nonlinear growth model as well as tracking of a maneuvering target. In the experiments, we also compare against approximate likelihood estimates obtained by particle filtering and extended Kalman filtering based methods. The experiments suggest that the higher-order unscented transforms may in some cases provide more accurate estimates.
\end{abstract}

\IEEEpeerreviewmaketitle

\section{Introduction}

\IEEEPARstart{T}{his} paper is an extended version of our article \cite{Kokkala+Solin+Sarkka:2014} where we considered parameter estimation in state-space models using expectation--maximization (EM) algorithms based on sigma-point and particle smoothers. In this paper, we extend our interest from EM algorithms to so called direct maximum likelihood based parameter estimation methods, where instead of using the EM algorithm, the marginal likelihood of the parameters is directly approximated using nonlinear filtering methods. In particular, we focus our interest to sigma-point filters which use high-order unscented Kalman filters and Gauss--Hermite Kalman filters to approximate the likelihood surface.

We consider state-space models of the following form:
\begin{equation}
\begin{split}\label{eq:dmodel}
  \vx_k &= \vf(\vx_{k-1},\vtheta) + \vq_{k-1}, \\
  \vy_k &= \vh(\vx_k,\vtheta) + \vr_k,
\end{split}
\end{equation}
where $\vx_k \in \R^n$ is the discrete-time state sequence with an initial distribution  $\vx_0 \sim \N(\vx_0 \mid \vm_0(\vtheta),\MP_0(\vtheta))$, $\vy_k \in \R^d$ is the measurement sequence, $\vq_k \sim \N(\vect{0},\MQ(\vtheta))$ is the Gaussian process noise sequence, $\vr_k \sim \N(\vect{0},\MR(\vtheta))$ is the Gaussian measurement error sequence, and $\vtheta \in \R^m$ is a static parameter vector. Typically, one is interested in computing the posterior distribution of the state $\vx_k$ given measurements up to time $k$, $p(\vx_k \mid \vy_1,\ldots,\vy_k)$, known as the filtering problem, or computing the posterior distribution of the state $\vx_k$ given all measurements, $p(\vx_k \mid \vy_1,\ldots,\vy_T)$, where $k\leq T$, known as the smoothing problem. In the general case, analytical expressions do not exist and we have to resort to approximative algorithms such as the sigma-point methods. See, for example, \cite{Sarkka:2013} for a general overview of Bayesian filtering and smoothing. 

While many filtering and smoothing algorithms are formulated assuming fixed static parameters $\vtheta$, in practice optimal values for these parameters are generally unknown. Therefore, methods for estimating the parameters from the data are desired. In this paper, we concentrate on maximum-likelihood methods, where the parameters are selected by maximizing the marginal likelihood, or equivalently the logarithm of the marginal likelihood, that is 
\begin{equation}
  \vtheta_\mathrm{ML}  = \mathrm{arg\,max}_{\vtheta} \log p(\vy_{1:T} \mid \vtheta).
\end{equation}
In linear systems with additive Gaussian noise, the likelihood can be evaluated using the Kalman filter \cite{Kalman:1960,Jazwinski:1970}. Many optimization algorithms utilize also the gradient of the log-likelihood. The gradient can be evaluated by so-called sensitivity equations, a recursion that is obtained by differentiating the Kalman filter recursion \cite{Gupta+Mehra:1974}. Alternatively, due to Fisher's identity, the gradient may be evaluated by differentiating an auxiliary function that can be computed during the smoothing pass \cite{Segal+Weinstein:1988,Olsson+Petersen+Lehn-Schioler:2007}. Instead of directly optimizing the likelihood, the expectation--maximization (EM) algorithm \cite{Dempster:1977} can be used to optimize parameters. The EM algorithm consists of iterating the expectation (E) step where a bound of the log-likelihood is computed using the current parameter estimates, and the maximization (M) step where the bound is maximized with respect to the parameters. The evaluation of the bound in the E-step is obtained by solving the smoothing problem. See \cite{Shumway+Stoffer:1982} for a discussion of applying the EM algorithm in state-space models. Note that in the linear-Gaussian case both gradient evaluation methods as well as the EM algorithm in principle converge to the same solution, namely, the parameter value that maximizes the log-likelihood.

In this paper, our interest lies in estimating the static parameters by maximum-likelihood estimation in the case of nonlinear state-space models with additive Gaussian noise, that is model~\eqref{eq:dmodel}. Formally, the marginal likelihood can be computed by marginalizing out the states from the joint distribution of the measurements and states using nonlinear filtering equations and the prediction error decomposition (see, \eg, \cite{Cappe+Moulines:2005,Sarkka:2013}), leading to similar methods as in the linear-Gaussian case. However, since the state variables $\vx$ cannot in general be marginalized out analytically, one needs to employ approximative methods. In the so called direct likelihood methods, the likelihood is approximated directly using approximative nonlinear filtering methods (see, \eg, \cite{Kitagawa:1987, Singer:2002, Poyiadjis+Doucet+Singh:2005, Cappe+Moulines:2005, Sarkka:2013}) and its maximum is found via nonlinear optimization. Similarly, the expectation--maximization (EM) algorithm can be employed, but the E-step cannot be solved exactly. Instead, the E-step is approximated with nonlinear smoothing algorithms (see, \eg, \cite{Roweis+Ghahramani:2001, Vaananen:2012, Gasperin+Juricic:2011, Schon+Wills+Ninness:2011, Kokkala+Solin+Sarkka:2014}).

The aim of this paper is to extend the results of our paper \cite{Kokkala+Solin+Sarkka:2014} by showing how high-order (\ie, third, fifth, seventh, and ninth order) unscented transforms and Gauss--Hermite integration based sigma-point methods can be used for approximate direct likelihood and EM-based parameter estimation in nonlinear state-space models. For EM, we also give closed-form expressions for the maximization step in a class of models that are linear in parameters and have additive noise.  We compare the unscented transform and Gauss--Hermite based sigma-point methods to linearization-based extended Kalman filter algorithms and Monte Carlo based particle filtering algorithms. We also provide an algorithm for computing the gradients required by the gradient-based optimization methods. Although we focus on maximum likelihood estimation, the provided algorithms can be easily extended to computation of maximum \textit{a~posteriori} estimates by including a prior distribution to the objective function.

\section{Sigma-Point Filtering and Smoothing}

Under our interpretation, sigma-point filtering and smoothing is derived by assuming Gaussian approximations for the state distributions, which enables the use of a Kalman filter like filtering recursion and a Rauch--Tung--Striebel backward pass for the smoothing distributions. The Gaussian filtering and smoothing equations contain expectations over Gaussian distributions which cannot be generally evaluated in closed form. The sigma-points arise from approximating these Gaussian integrals by weighted sums determined by some cubature (multi-dimensional quadrature) formula. Hence, we interpret the different sigma-point methods as incarnations of different integral approximations.

In the following, we first present the assumed Gaussian density filtering and smoothing framework. Then, we discuss various different cubature rules for approximating the Gaussian integrals. Finally, we show how the cubature rules are applied to the assumed Gaussian density filtering and smoothing framework to obtain the filtering and smoothing equations explicitly in the sigma-point form.

\subsection{General Gaussian Filtering and Smoothing}
\label{sec:gaussianfiltering}
Assumed density Gaussian filtering (see \cite{Ito+Xiong:2000, Sarkka+Hartikainen:2010, Sarkka:2013}) is based on assuming that the filtering distributions are approximately Gaussian, that is, assuming means $\vm_{k \mid k}$ and covariances $\MP_{k \mid k}$ such that
\begin{equation}
  p(\vx_{k} \mid \vy_{1:k}) \approx \N(\vx_{k} \mid \vm_{k \mid k}, \MP_{k \mid k})
\end{equation}
as well as means $\vm_{k \mid k+1}$ and covariances $\MP_{k \mid k+1}$ such that
\begin{equation}
  p(\vx_{k+1} \mid \vy_{1:k}) \approx \N(\vx_{k+1} \mid \vm_{k \mid k+1}, \MP_{k \mid k+1}).
\end{equation}
The filtering equations of the resulting Gaussian filter \cite{Wu+Hu+Wu+Hu:2006, Ito+Xiong:2000} consist of a \emph{prediction step} and an \emph{update step}. In the prediction step, we compute the state mean and covariance of the distribution $p(\vx_k \mid \vy_{1:k-1})$ using the Gaussian approximation for $p(\vx_{k-1} \mid \vy_{1:k-1})$. The resulting equations are
\begin{equation}
  \begin{split} \label{eq:adfpred}
     \vm_{k \mid k-1} &= \E[\vf(\vx_{k-1})], \\
     \MP_{k \mid k-1} &= \E[(\vf(\vx_{k-1}) - \vm_{k \mid k-1})  \\
                      & \qquad \times (\vf(\vx_{k-1}) - \vm_{k \mid k-1})^\T] + \MQ,
  \end{split}
\end{equation}
where the expectations are taken with respect to the distribution $\vx_{k-1} \sim \N(\vm_{k-1 \mid k-1},\MP_{k-1 \mid k-1})$.

In the corresponding update step, we assume a Gaussian density $p(\vx_k \mid \vy_{1:k-1}) = \N(\vx_k \mid \vm_{k \mid k-1} , \MP_{k \mid k-1})$ and compute the state mean and covariance for the distribution $p(\vx_k \mid \vy_{1:k})$. The resulting equations are 
\begin{equation}
  \begin{split} \label{eq:adfupda} 
          \vmu_{k} &= \E[\vh(\vx_{k})], \\
           \MS_{k} &= \E[(\vh(\vx_{k}) - \vmu_{k}) \, 
                         (\vh(\vx_{k}) - \vmu_{k})^\T] + \MR, \\
             \MC_k &= \E[(\vx_k - \vm_{k \mid k-1}) \,
                         (\vh(\vx_k) - \vmu_k)^\T], \\
           \MK_{k} &= \MC_k \, \MS_{k}^{-1}, \\
    \vm_{k \mid k} &= \vm_{k \mid k-1} + \MK_{k} \, (\vy_k - \vmu_k), \\
    \MP_{k \mid k} &= \MP_{k \mid k-1} - \MK_{k} \, \MS_{k} \, \MK_{k}^\T,
  \end{split}
\end{equation}
where the expectations are taken with respect to the distribution $\vx_k \sim \N(\vm_{k \mid k-1}, \MP_{k \mid k-1})$. 

The smoothing distributions $p(\vx_k \mid \vy_{1:T})$ are obtained from a backward pass, that is, starting from $k=T$ and iterating backwards in time. On each step, the smoothing density of $\vx_{k+1}$ is assumed to be Gaussian: $p(\vx_{k+1} \mid \vy_{1:T}) = \N(\vx_{k+1} \mid \vm_{k+1 \mid T}, \MP_{k+1 \mid T})$. The mean and covariance for $p(\vx_{k} \mid \vy_{1:T})$ are then computed from the previous Gaussian smoothing density and the Gaussian filtering densities using the Rauch--Tung--Striebel backward pass \cite{Rauch+Tung+Striebel:1965,Gelb:1974} as follows \cite{Sarkka+Hartikainen:2010}:  
\begin{equation}
  \begin{split} \label{eq:adfrts}
     \vm_{k+1 \mid k} &= \E[\vf(\vx_{k})], \\
     \MP_{k+1 \mid k} &= \E[(\vf(\vx_{k}) - \vm_{k+1 \mid k}) \\
                      & \qquad \times (\vf(\vx_{k}) - \vm_{k+1 \mid k})^\T] + \MQ, \\
            \MD_{k+1} &= \E[(\vx_{k} - \vm_{k \mid k}) \,
                            (\vf(\vx_{k}) - \vm_{k+1 \mid k})^\T], \\
              \MG_{k} &= \MD_{k+1} \, [\MP_{k+1 \mid k}]^{-1}, \\
       \vm_{k \mid T} &= \vm_{k \mid k} + \MG_{k} \, (\vm_{k+1 \mid T} - \vm_{k+1 \mid k}), \\
       \MP_{k \mid T} &= \MP_{k \mid k} + \MG_{k} \, (\MP_{k+1 \mid T} - \MP_{k+1 \mid k}) \, \MG_{k}^\T,
  \end{split}
\end{equation}
where the expectations are taken with respect to the distribution $\vx_k \sim \N(\vm_{k \mid k}, \MP_{k \mid k})$. The pairwise joint smoothing distributions $p(\vx_k, \vx_{k-1} \mid \vy_{1:T})$ are also of interest since they are used in the expectation--maximization algorithm (see Section~\ref{sec:em}). Gaussian approximations for these distributions are obtained as a by-product of the smoothing backward pass results as follows (see, \eg, \cite[p.~189]{Sarkka:2013}).
\begin{multline} \label{eq:gaussian-pairwise}
  p(\vx_k, \vx_{k-1} \mid \vy_{1:T}) \approx \vphantom{\bigg|} \\
  \N\!\left( 
      \begin{pmatrix} \vx_k \\ \vx_{k-1} \end{pmatrix} \bigg |
      \begin{pmatrix} \vm_{k \mid T} \\ \vm_{k-1 \mid T} \end{pmatrix}, 
      \begin{pmatrix} 
        \MP_{k \mid T}          & \MP_{k \mid T} \MG_{k-1}^\T \\
        \MG_{k-1}\MP_{k \mid T} & \MP_{k-1 \mid T}
      \end{pmatrix} \right).
\end{multline}

\subsection{Approximating the Gaussian Integrals}
\label{sec:sigmapointschemes}
As we saw in the previous section, during the evaluation of the prediction and update steps of the Gaussian filter and smoother, we need to solve a set of Gaussian integrals on each step. These integrals are of the following form:
\begin{equation} \label{eq:qint}
  \E[\vg(\vx)] = \int_{\R^n} \vg(\vx) \, \N(\vx \mid \vm, \MP) \dd \vx,
\end{equation}
where $\vg: \R^n \to \R^d$ is the integrand and the weighting function $\N(\vx \mid \vm, \MP)$ is a multi-dimensional Gaussian density with mean $\vm$ and covariance matrix $\MP$. In this paper, these integrals are computed by using multi-dimensional generalizations of Gaussian quadratures---also referred to as Gaussian \emph{cubatures} \cite{Cools:1997}. They give approximations of the form
\begin{equation}
  \E[\vg(\vx)] \approx \sum_i w_i \, \vg(\vx_i),
\end{equation}
where the weights $w_i$ and sigma-points $\vx_i$ are functions of the mean $\vm$ and covariance $\MP$ of the Gaussian weighting function. The sigma-points are positioned as follows:
\begin{equation}
  \vx_i = \vm + \ML \, \vxi_i,
\end{equation}
where $\vxi_i$ are method specific unit sigma-points, and $\ML$ is a matrix square-root factor such that $\MP = \ML\,\ML^\T$ (\eg, the Cholesky decomposition of $\MP$). The differences in the methods come from different choices of weights and unit sigma-points.

In the following we briefly introduce a number of schemes for choosing the weights and sigma-points. The difference between these schemes stems from a trade-off between the number of sigma-points (required function evaluations) versus precision in the approximation. The degree of approximation is quantified by the highest polynomial order, $p$, for which the method is exact.

\begin{figure*}[!t]
  \pgfplotsset{
    compat=newest,    
    clip=false,
    hide y axis,
    tick align=outside,
    minor tick num=1,
  }
  \setlength{\figurewidth}{0.15\textwidth}
  \setlength{\figureheight}{0.15\textwidth}
  \centering \sffamily \scriptsize
  %
  %\tikzsetnextfilename{fig01a}
  \subfloat[Symmetric (order~3)]{% This file was created by matlab2tikz v0.4.2.
% Copyright (c) 2008--2013, Nico Schlömer <nico.schloemer@gmail.com>
% All rights reserved.
% 
% The latest updates can be retrieved from
%   http://www.mathworks.com/matlabcentral/fileexchange/22022-matlab2tikz
% where you can also make suggestions and rate matlab2tikz.
% 
% 
% 
\begin{tikzpicture}

\begin{axis}[%
width=\figurewidth,
height=\figureheight,
scale only axis,
xmin=-5,
xmax=5,
xtick={-4,  0,  4},
ymin=-5,
ymax=5,
axis x line*=bottom,
axis y line*=left
]
\addplot [
color=blue,
mark size=0.0pt,
only marks,
mark=*,
mark options={solid,fill=black,draw=black},
forget plot
]
table[row sep=crcr]{
0 0\\
};
\addplot [
color=blue,
mark size=2.2pt,
only marks,
mark=*,
mark options={solid,fill=black,draw=black},
forget plot
]
table[row sep=crcr]{
1.4142135623731 0\\
};
\addplot [
color=blue,
mark size=2.2pt,
only marks,
mark=*,
mark options={solid,fill=black,draw=black},
forget plot
]
table[row sep=crcr]{
0 1.4142135623731\\
};
\addplot [
color=blue,
mark size=2.2pt,
only marks,
mark=*,
mark options={solid,fill=black,draw=black},
forget plot
]
table[row sep=crcr]{
-1.4142135623731 -0\\
};
\addplot [
color=blue,
mark size=2.2pt,
only marks,
mark=*,
mark options={solid,fill=black,draw=black},
forget plot
]
table[row sep=crcr]{
-0 -1.4142135623731\\
};
\end{axis}
\end{tikzpicture}%
}%
  \hfil
  %\tikzsetnextfilename{fig01b}
  \subfloat[Symmetric (order~5)]{% This file was created by matlab2tikz v0.4.2.
% Copyright (c) 2008--2013, Nico Schlömer <nico.schloemer@gmail.com>
% All rights reserved.
% 
% The latest updates can be retrieved from
%   http://www.mathworks.com/matlabcentral/fileexchange/22022-matlab2tikz
% where you can also make suggestions and rate matlab2tikz.
% 
% 
% 
\begin{tikzpicture}

\begin{axis}[%
width=\figurewidth,
height=\figureheight,
scale only axis,
xmin=-5,
xmax=5,
xtick={-4,  0,  4},
ymin=-5,
ymax=5,
axis x line*=bottom,
axis y line*=left
]
\addplot [
color=blue,
mark size=3.0pt,
only marks,
mark=*,
mark options={solid,fill=black,draw=black},
forget plot
]
table[row sep=crcr]{
0 0\\
};
\addplot [
color=blue,
mark size=1.5pt,
only marks,
mark=*,
mark options={solid,fill=black,draw=black},
forget plot
]
table[row sep=crcr]{
1.73205080756888 0\\
};
\addplot [
color=blue,
mark size=1.5pt,
only marks,
mark=*,
mark options={solid,fill=black,draw=black},
forget plot
]
table[row sep=crcr]{
-1.73205080756888 -0\\
};
\addplot [
color=blue,
mark size=1.5pt,
only marks,
mark=*,
mark options={solid,fill=black,draw=black},
forget plot
]
table[row sep=crcr]{
0 1.73205080756888\\
};
\addplot [
color=blue,
mark size=1.5pt,
only marks,
mark=*,
mark options={solid,fill=black,draw=black},
forget plot
]
table[row sep=crcr]{
-0 -1.73205080756888\\
};
\addplot [
color=blue,
mark size=0.7pt,
only marks,
mark=*,
mark options={solid,fill=black,draw=black},
forget plot
]
table[row sep=crcr]{
1.73205080756888 1.73205080756888\\
};
\addplot [
color=blue,
mark size=0.7pt,
only marks,
mark=*,
mark options={solid,fill=black,draw=black},
forget plot
]
table[row sep=crcr]{
-1.73205080756888 -1.73205080756888\\
};
\addplot [
color=blue,
mark size=0.7pt,
only marks,
mark=*,
mark options={solid,fill=black,draw=black},
forget plot
]
table[row sep=crcr]{
1.73205080756888 -1.73205080756888\\
};
\addplot [
color=blue,
mark size=0.7pt,
only marks,
mark=*,
mark options={solid,fill=black,draw=black},
forget plot
]
table[row sep=crcr]{
-1.73205080756888 1.73205080756888\\
};
\end{axis}
\end{tikzpicture}%
}%
  \hfil
  %\tikzsetnextfilename{fig01c}
  \subfloat[Symmetric (order~7)]{% This file was created by matlab2tikz v0.4.2.
% Copyright (c) 2008--2013, Nico Schlömer <nico.schloemer@gmail.com>
% All rights reserved.
% 
% The latest updates can be retrieved from
%   http://www.mathworks.com/matlabcentral/fileexchange/22022-matlab2tikz
% where you can also make suggestions and rate matlab2tikz.
% 
% 
% 
\begin{tikzpicture}

\begin{axis}[%
width=\figurewidth,
height=\figureheight,
scale only axis,
xmin=-5,
xmax=5,
xtick={-4,  0,  4},
ymin=-5,
ymax=5,
axis x line*=bottom,
axis y line*=left
]
\addplot [
color=blue,
mark size=3.7pt,
only marks,
mark=*,
mark options={solid,fill=black,draw=black},
forget plot
]
table[row sep=crcr]{
0 0\\
};
\addplot [
color=blue,
mark size=0.9pt,
only marks,
mark=*,
mark options={solid,fill=black,draw=black},
forget plot
]
table[row sep=crcr]{
2.33441421833898 0\\
};
\addplot [
color=blue,
mark size=0.9pt,
only marks,
mark=*,
mark options={solid,fill=black,draw=black},
forget plot
]
table[row sep=crcr]{
-2.33441421833898 -0\\
};
\addplot [
color=blue,
mark size=0.9pt,
only marks,
mark=*,
mark options={solid,fill=black,draw=black},
forget plot
]
table[row sep=crcr]{
0 2.33441421833898\\
};
\addplot [
color=blue,
mark size=0.9pt,
only marks,
mark=*,
mark options={solid,fill=black,draw=black},
forget plot
]
table[row sep=crcr]{
-0 -2.33441421833898\\
};
\addplot [
color=blue,
mark size=2.7pt,
only marks,
mark=*,
mark options={solid,fill=white,draw=black},
forget plot
]
table[row sep=crcr]{
0.741963784302726 0\\
};
\addplot [
color=blue,
mark size=2.7pt,
only marks,
mark=*,
mark options={solid,fill=white,draw=black},
forget plot
]
table[row sep=crcr]{
-0.741963784302726 -0\\
};
\addplot [
color=blue,
mark size=2.7pt,
only marks,
mark=*,
mark options={solid,fill=white,draw=black},
forget plot
]
table[row sep=crcr]{
0 0.741963784302726\\
};
\addplot [
color=blue,
mark size=2.7pt,
only marks,
mark=*,
mark options={solid,fill=white,draw=black},
forget plot
]
table[row sep=crcr]{
-0 -0.741963784302726\\
};
\addplot [
color=blue,
mark size=0.3pt,
only marks,
mark=*,
mark options={solid,fill=black,draw=black},
forget plot
]
table[row sep=crcr]{
2.33441421833898 2.33441421833898\\
};
\addplot [
color=blue,
mark size=0.3pt,
only marks,
mark=*,
mark options={solid,fill=black,draw=black},
forget plot
]
table[row sep=crcr]{
-2.33441421833898 -2.33441421833898\\
};
\addplot [
color=blue,
mark size=0.3pt,
only marks,
mark=*,
mark options={solid,fill=black,draw=black},
forget plot
]
table[row sep=crcr]{
2.33441421833898 -2.33441421833898\\
};
\addplot [
color=blue,
mark size=0.3pt,
only marks,
mark=*,
mark options={solid,fill=black,draw=black},
forget plot
]
table[row sep=crcr]{
-2.33441421833898 2.33441421833898\\
};
\addplot [
color=blue,
mark size=2.9pt,
only marks,
mark=*,
mark options={solid,fill=black,draw=black},
forget plot
]
table[row sep=crcr]{
0.741963784302726 0.741963784302726\\
};
\addplot [
color=blue,
mark size=2.9pt,
only marks,
mark=*,
mark options={solid,fill=black,draw=black},
forget plot
]
table[row sep=crcr]{
-0.741963784302726 -0.741963784302726\\
};
\addplot [
color=blue,
mark size=2.9pt,
only marks,
mark=*,
mark options={solid,fill=black,draw=black},
forget plot
]
table[row sep=crcr]{
0.741963784302726 -0.741963784302726\\
};
\addplot [
color=blue,
mark size=2.9pt,
only marks,
mark=*,
mark options={solid,fill=black,draw=black},
forget plot
]
table[row sep=crcr]{
-0.741963784302726 0.741963784302726\\
};
\end{axis}
\end{tikzpicture}%
}%
  \hfil
  %\tikzsetnextfilename{fig01d}
  \subfloat[Symmetric (order~9)]{% This file was created by matlab2tikz v0.4.2.
% Copyright (c) 2008--2013, Nico Schlömer <nico.schloemer@gmail.com>
% All rights reserved.
% 
% The latest updates can be retrieved from
%   http://www.mathworks.com/matlabcentral/fileexchange/22022-matlab2tikz
% where you can also make suggestions and rate matlab2tikz.
% 
% 
% 
\begin{tikzpicture}

\begin{axis}[%
width=\figurewidth,
height=\figureheight,
scale only axis,
xmin=-5,
xmax=5,
xtick={-4,  0,  4},
ymin=-5,
ymax=5,
axis x line*=bottom,
axis y line*=left
]
\addplot [
color=blue,
mark size=2.4pt,
only marks,
mark=*,
mark options={solid,fill=black,draw=black},
forget plot
]
table[row sep=crcr]{
0 0\\
};
\addplot [
color=blue,
mark size=0.3pt,
only marks,
mark=*,
mark options={solid,fill=black,draw=black},
forget plot
]
table[row sep=crcr]{
2.8569700138728 0\\
};
\addplot [
color=blue,
mark size=0.3pt,
only marks,
mark=*,
mark options={solid,fill=black,draw=black},
forget plot
]
table[row sep=crcr]{
-2.8569700138728 -0\\
};
\addplot [
color=blue,
mark size=0.3pt,
only marks,
mark=*,
mark options={solid,fill=black,draw=black},
forget plot
]
table[row sep=crcr]{
0 2.8569700138728\\
};
\addplot [
color=blue,
mark size=0.3pt,
only marks,
mark=*,
mark options={solid,fill=black,draw=black},
forget plot
]
table[row sep=crcr]{
-0 -2.8569700138728\\
};
\addplot [
color=blue,
mark size=1.5pt,
only marks,
mark=*,
mark options={solid,fill=black,draw=black},
forget plot
]
table[row sep=crcr]{
1.35562617997427 0\\
};
\addplot [
color=blue,
mark size=1.5pt,
only marks,
mark=*,
mark options={solid,fill=black,draw=black},
forget plot
]
table[row sep=crcr]{
-1.35562617997427 -0\\
};
\addplot [
color=blue,
mark size=1.5pt,
only marks,
mark=*,
mark options={solid,fill=black,draw=black},
forget plot
]
table[row sep=crcr]{
0 1.35562617997427\\
};
\addplot [
color=blue,
mark size=1.5pt,
only marks,
mark=*,
mark options={solid,fill=black,draw=black},
forget plot
]
table[row sep=crcr]{
-0 -1.35562617997427\\
};
\addplot [
color=blue,
mark size=0.1pt,
only marks,
mark=*,
mark options={solid,fill=black,draw=black},
forget plot
]
table[row sep=crcr]{
2.8569700138728 2.8569700138728\\
};
\addplot [
color=blue,
mark size=0.1pt,
only marks,
mark=*,
mark options={solid,fill=black,draw=black},
forget plot
]
table[row sep=crcr]{
-2.8569700138728 -2.8569700138728\\
};
\addplot [
color=blue,
mark size=0.1pt,
only marks,
mark=*,
mark options={solid,fill=black,draw=black},
forget plot
]
table[row sep=crcr]{
2.8569700138728 -2.8569700138728\\
};
\addplot [
color=blue,
mark size=0.1pt,
only marks,
mark=*,
mark options={solid,fill=black,draw=black},
forget plot
]
table[row sep=crcr]{
-2.8569700138728 2.8569700138728\\
};
\addplot [
color=blue,
mark size=0.2pt,
only marks,
mark=*,
mark options={solid,fill=black,draw=black},
forget plot
]
table[row sep=crcr]{
2.8569700138728 1.35562617997427\\
};
\addplot [
color=blue,
mark size=0.2pt,
only marks,
mark=*,
mark options={solid,fill=black,draw=black},
forget plot
]
table[row sep=crcr]{
-2.8569700138728 -1.35562617997427\\
};
\addplot [
color=blue,
mark size=0.2pt,
only marks,
mark=*,
mark options={solid,fill=black,draw=black},
forget plot
]
table[row sep=crcr]{
2.8569700138728 -1.35562617997427\\
};
\addplot [
color=blue,
mark size=0.2pt,
only marks,
mark=*,
mark options={solid,fill=black,draw=black},
forget plot
]
table[row sep=crcr]{
-2.8569700138728 1.35562617997427\\
};
\addplot [
color=blue,
mark size=0.2pt,
only marks,
mark=*,
mark options={solid,fill=black,draw=black},
forget plot
]
table[row sep=crcr]{
1.35562617997427 2.8569700138728\\
};
\addplot [
color=blue,
mark size=0.2pt,
only marks,
mark=*,
mark options={solid,fill=black,draw=black},
forget plot
]
table[row sep=crcr]{
-1.35562617997427 -2.8569700138728\\
};
\addplot [
color=blue,
mark size=0.2pt,
only marks,
mark=*,
mark options={solid,fill=black,draw=black},
forget plot
]
table[row sep=crcr]{
-1.35562617997427 2.8569700138728\\
};
\addplot [
color=blue,
mark size=0.2pt,
only marks,
mark=*,
mark options={solid,fill=black,draw=black},
forget plot
]
table[row sep=crcr]{
1.35562617997427 -2.8569700138728\\
};
\addplot [
color=blue,
mark size=1.0pt,
only marks,
mark=*,
mark options={solid,fill=black,draw=black},
forget plot
]
table[row sep=crcr]{
1.35562617997427 1.35562617997427\\
};
\addplot [
color=blue,
mark size=1.0pt,
only marks,
mark=*,
mark options={solid,fill=black,draw=black},
forget plot
]
table[row sep=crcr]{
-1.35562617997427 -1.35562617997427\\
};
\addplot [
color=blue,
mark size=1.0pt,
only marks,
mark=*,
mark options={solid,fill=black,draw=black},
forget plot
]
table[row sep=crcr]{
1.35562617997427 -1.35562617997427\\
};
\addplot [
color=blue,
mark size=1.0pt,
only marks,
mark=*,
mark options={solid,fill=black,draw=black},
forget plot
]
table[row sep=crcr]{
-1.35562617997427 1.35562617997427\\
};
\end{axis}
\end{tikzpicture}%
}%
  \hfil
  %\tikzsetnextfilename{fig01e}
  \subfloat[Gauss--Hermite]{\input{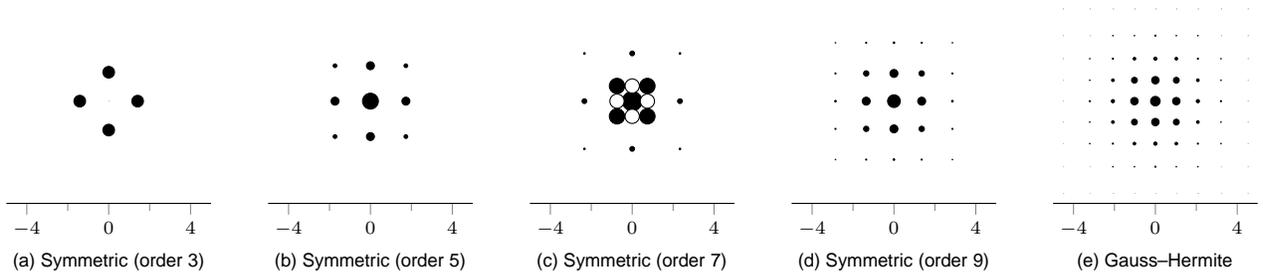}}%
  \caption{Unit sigma-points in two dimensions for each of the methods. The absolute value of the weights are indicated by the point size, positive weights being black, negative weights white. (a)~The symmetric cubature rule of order $p=3$ with 4~points. (b) The symmetric cubature rule of order $p=5$ with 9~points. (c)~The symmetric cubature rule of order $p=7$ with 17~points. (d)~The symmetric cubature rule of order $p=9$ with 25 points. (e)~For comparison, the Gauss--Hermite (order $p=9$) sigma-points (81~points, many of which with very small weights) are also shown.}
  \label{fig:sigma-points}
\end{figure*}

\begin{LaTeXdescription}
\item[Unscented transform.] The \emph{unscented transform} (UT, \cite{Julier+Uhlmann+Durrant-Whyte:1995, Julier+Uhlmann+Durrant-Whyte:2000}) uses a set of $2n+1$ cubature points located in the center and on the surface of an $n$-sphere. The radius and the weights can be controlled using a set of parameters. The cubature points are given by:
\begin{equation*}
\begin{split}
  \vxi_0 &= \vect{0}, \\
  \vxi_i &= \left\{ \begin{array}{ll}
    \hphantom{-}\sqrt{\lambda + n} \, \ve_i,     & i = 1,\ldots,n, \\
               -\sqrt{\lambda + n} \, \ve_{i-n}, & i = n+1,\ldots,2n, \\
  \end{array} \right.
\end{split}
\label{eq:uxi}
\end{equation*}
where $\ve_i$ denotes a unit vector to the direction of coordinate axis $i$, and the weights are defined as follows:
\begin{equation*}
\begin{split} \label{eq:uw}
  w^{(0)} &= \left\{ \begin{array}{ll}
    \frac{\lambda}{n + \lambda}, &
      \text{for mean terms}, \\
    \frac{\lambda}{n + \lambda} + (1 - \alpha^2 + \beta), &
      \text{for covariance terms},
  \end{array} \right. \\
  w^{(i)} &= \frac{1}{2 (n + \lambda)}, \qquad i = 1,\ldots,2n,
\end{split}
\end{equation*}
where $\lambda = \alpha^2 (n + \kappa) - n$ and $\alpha$, $\beta$, and $\kappa$ are parameters of the method.
  \item[Symmetric, 3$^\text{rd}$ order.] A widely applicable sigma-point scheme is constructed by setting the unscented transform parameters to $\alpha=\pm1$, $\beta=0$, and $\kappa=0$ \cite{Wu+Hu+Wu+Hu:2006}. This is also known as the $3^\text{rd}$ order symmetric spherical--radial cubature method (CKF, \cite{Arasaratnam+Haykin+Hurd:2010}; see \cite{Solin:2010} for the explicit connection). This method utilizes a scaled and rotated set of $2n$ points, which are selected to be at the intersections of an $n$-sphere and the coordinate axes:
\begin{equation*} \label{eq:cxi}
  \vxi_i = \left\{ \begin{array}{ll}
    \hphantom{-} \sqrt{n} \, \ve_i,     & i = 1,2,\ldots,n, \\
                -\sqrt{n} \, \ve_{i-n}, & i = n+1,\ldots,2n. \\
  \end{array} \right.
\end{equation*}
The weights are defined as $w_i = 1/(2n)$ for $i = 1,2,\ldots,2n$. The number of evaluation points is a linear function of the state dimension. The corresponding sigma-point filter is referred to as UKF~3.
  \item[Symmetric, 5$^\text{th}$ order.] Building upon the work of McNamee and Stenger \cite{McNamee+Stenger:1967}, it is possible to find explicit fully symmetric integration formulas of higher order than three. These integration schemes are exact for symmetric polynomials up to a given order $p$. For order $p=5$, the number of required sigma-points is $2n^2+1$. The corresponding sigma-point filter is referred to as UKF~5.
  \item[Symmetric, 7$^\text{th}$ order.] For order $p=7$, the number of required sigma-points is $\frac{1}{3}(4n^3+8n+3)$, meaning that they scale cubicly with the number of state dimensions. The corresponding sigma-point filter is referred to as UKF~7.

  \item[Symmetric, 9$^\text{th}$ order.] For order $p=9$, the number of required sigma-points is $\frac{1}{3}(2n^4-4n^3+22n^2-8n+3)$. The corresponding sigma-point filter is referred to as UKF~9. If required, even higher order methods can be constructed in the spirit of \cite{McNamee+Stenger:1967}.
  \item[Gauss--Hermite.] The $n$-dimensional Gauss--Hermite quadrature method forms the sigma-points as a Cartesian product of the one-dimensional Gauss--Hermite quadratures, and the weights are simply products of the one-dimensional weights \cite{Cools:1997, Ito+Xiong:2000, Wu+Hu+Wu+Hu:2006}. The disadvantage of this method is that with a $p$th order GH approximation (exact for polynomials up to order $p$), the required number of evaluation points is $p^n$, the number growing exponentially with state dimension $n$. The corresponding filter is referred to as GHKF.
\end{LaTeXdescription}

The exact formulas for the higher-order methods become lengthy and have been omitted here for brevity (see, \cite{McNamee+Stenger:1967,Lerner:2002,Wu+Hu+Wu+Hu:2006}, for implementation details and discussion). Figure~\ref{fig:sigma-points} gives a pictorial example of how the points and weights are placed in two dimensions ($n=2$) for each of the methods. Note that even though the 5$^\text{th}$ and 9$^\text{th}$ order methods do not have negative weights when $n=2$, they have negative weights in other dimensions.

\begin{figure}[!t]
  \pgfplotsset{
    compat=newest,    
    clip=true,
    tick align=inside,
    minor tick num=1,
  }
  %
  %\tikzsetnextfilename{fig02}
  \setlength{\figurewidth}{0.8\columnwidth}
  \setlength{\figureheight}{0.6\columnwidth}
  \centering \sffamily \scriptsize
  % This file was created by matlab2tikz v0.4.2.
% Copyright (c) 2008--2013, Nico Schlömer <nico.schloemer@gmail.com>
% All rights reserved.
% 
% The latest updates can be retrieved from
%   http://www.mathworks.com/matlabcentral/fileexchange/22022-matlab2tikz
% where you can also make suggestions and rate matlab2tikz.
% 
% 
% 
%
% defining custom colors
\definecolor{mycolor1}{rgb}{0,1,1}%
\definecolor{mycolor1EKF}{rgb}{0.00000,0.44700,0.74100}%
\definecolor{mycolor2UKF3}{rgb}{0.85000,0.32500,0.09800}%
\definecolor{mycolor3UKF5}{rgb}{0.92900,0.69400,0.12500}%
\definecolor{mycolor4UKF7}{rgb}{0.49400,0.18400,0.55600}%
\definecolor{mycolor5UKF9}{rgb}{0.30100,0.74500,0.93300}%
\begin{tikzpicture}

\begin{axis}[%
width=\figurewidth,
height=\figureheight,
scale only axis,
xmin=1,
xmax=9,
xlabel={State dimension, $n$},
ymode=log,
ymin=1,
ymax=1000000000,
yminorticks=true,
ylabel={Number of sigma-points},
legend style={at={(0.03,0.97)},anchor=north west,draw=black,fill=white,legend cell align=left}
]
\addplot [
color=mycolor2UKF3,
solid
]
table[row sep=crcr]{
1 3\\
2 5\\
3 7\\
4 9\\
5 11\\
6 13\\
7 15\\
8 17\\
9 19\\
};
\addlegendentry{Symmetric ($p=3$)};

\addplot [
color=mycolor3UKF5,
solid
]
table[row sep=crcr]{
1 3\\
2 9\\
3 19\\
4 33\\
5 51\\
6 73\\
7 99\\
8 129\\
9 163\\
};
\addlegendentry{Symmetric ($p=5$)};

\addplot [
color=mycolor4UKF7,
solid
]
table[row sep=crcr]{
1 5\\
2 17\\
3 45\\
4 97\\
5 181\\
6 305\\
7 477\\
8 705\\
9 997\\
};
\addlegendentry{Symmetric ($p=7$)};

\addplot [
color=mycolor5UKF9,
solid
]
table[row sep=crcr]{
1 5\\
2 25\\
3 77\\
4 193\\
5 421\\
6 825\\
7 1485\\
8 2497\\
9 3973\\
};
\addlegendentry{Symmetric ($p=9$)};

\addplot [
color=black,
dashed
]
table[row sep=crcr]{
1 3\\
2 9\\
3 27\\
4 81\\
5 243\\
6 729\\
7 2187\\
8 6561\\
9 19683\\
};
\addlegendentry{Gauss--Hermite};

\addplot [
color=black,
dashed,
forget plot
]
table[row sep=crcr]{
1 5\\
2 25\\
3 125\\
4 625\\
5 3125\\
6 15625\\
7 78125\\
8 390625\\
9 1953125\\
};
\addplot [
color=black,
dashed,
forget plot
]
table[row sep=crcr]{
1 7\\
2 49\\
3 343\\
4 2401\\
5 16807\\
6 117649\\
7 823543\\
8 5764801\\
9 40353607\\
};
\addplot [
color=black,
dashed,
forget plot
]
table[row sep=crcr]{
1 9\\
2 81\\
3 729\\
4 6561\\
5 59049\\
6 531441\\
7 4782969\\
8 43046721\\
9 387420489\\
};
\end{axis}
\end{tikzpicture}%
  \caption{Scaling of the number of sigma-points for each of the symmetric methods (solid lines). The required number of sigma-points for the Gauss--Hermite cubature of corresponding order $p$ is visualized by the dashed lines.}
  \label{fig:number-of-sigma-points}
  \vspace*{-12pt} %
\end{figure}
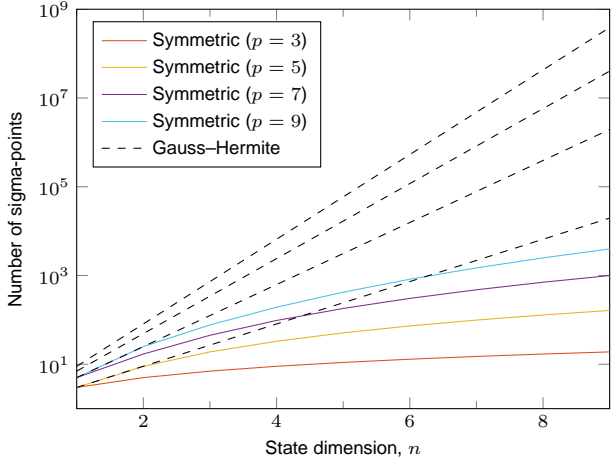

For higher state dimensions, Figure~\ref{fig:number-of-sigma-points} shows how the number of required points scale in each of the schemes. The exponentially growing number of evaluation points for Gauss--Hermite is apparent in Figure~\ref{fig:number-of-sigma-points}. In the UKFs, the number of evaluation points grow polynomially. McNamee and Stenger provide the following bound for the number of evaluation points for the fully symmetric integration formulas of arbitrary degree $p=2k+1$ in $n$-space: $\mathrm{O}((2n)^k/k!)$. Note that while in this paper we focus on the higher-order methods based on McNamee and Stenger, alternative cubature rules have also been suggested (see \cite{Crouse:2014-3D,Jia+Xin+Cheng:2013}).

\subsection{Sigma-Point Filtering and Smoothing}
\label{sec:sigmapointfiltering}
The following sigma-point filtering and smoothing equations are obtained by selecting a cubature rule, for example, one of the rules discussed in Section~\ref{sec:sigmapointschemes} and substituting it in place of the expectations in the Gaussian filtering and smoothing equations (Sec.~\ref{sec:gaussianfiltering}, Eqs.~\ref{eq:adfpred}--\ref{eq:adfrts}).

In the following equations, we denote the lower triangular matrix square-root (Cholesky) factor of a covariance matrix $\MP$ by $\ML$ so that for example $\MP_{k \mid k-1} = \ML_{k \mid k-1} \ML_{k \mid k-1}^\T$. The prediction step is 
\begin{equation} \label{eq:spfpred}
  \begin{split} 
     \vm_{k \mid k-1} &=  \sum_i w_i\,\vf\left(\vm_{k-1 \mid k-1} + \ML_{k-1 \mid k-1}\,\vxi_i \right), \\
     \MP_{k \mid k-1} &= \sum_i \Big\{ w_i\, \left(\vf\left(\vm_{k-1 \mid k-1} + \ML_{k-1 \mid k-1}\,\vxi_i\right) - \vm_{k \mid k-1} \right) \\
                      &\times \left( \vf\left(\vm_{k-1 \mid k-1} + \ML_{k-1 \mid k-1}\,\vxi_i \right) - \vm_{k \mid k-1}\right)^\T \Big\} + \MQ
  \end{split}
\end{equation}
and the update step is 
\begin{equation} \label{eq:spfupda} 
  \begin{split} 
          \vmu_{k} &= \sum_i w_i\,\vh\left(\vm_{k \mid k-1} + \ML_{k \mid k-1}\,\vxi_i\right), \\
           \MS_{k} &= \sum_i \Big\{ w_i\, \left( \vh\left(\vm_{k \mid k-1} + \ML_{k \mid k-1}\,\vxi_i\right) - \vmu_k \right) \\
                   &\times \left(\vh\left(\vm_{k \mid k-1} + \ML_{k \mid k-1}\,\vxi_i\right) - \mu_k \right)^\T \Big\} + \MR, \\
          \MC_k    &= \sum_i \Big\{ w_i\,  \ML_{k \mid k-1}\,\vxi_i \\ 
                   &\times \left(\vh\left(\vm_{k \mid k-1} + \ML_{k \mid k-1}\,\vxi_i \right) - \vmu_k \right)^\T \Big\}, \\                 
           \MK_{k} &= \MC_k \, \MS_{k}^{-1}, \\
    \vm_{k \mid k} &= \vm_{k \mid k-1} + \MK_{k} \, (\vy_k - \vmu_k), \\
    \MP_{k \mid k} &= \MP_{k \mid k-1} - \MK_{k} \, \MS_{k} \, \MK_{k}^\T.
  \end{split}
\end{equation}
The Rauch--Tung--Striebel smoother equations are 
\begin{equation} \label{eq:spfrts}
  \begin{split} 
     \vm_{k+1 \mid k} &=  \sum_i \left\{ w_i\,\vf\left(\vm_{k \mid k} + \ML_{k \mid k}\,\vxi_i \right) \right\}, \\
     \MP_{k+1 \mid k} &= \sum_i \Big\{ w_i\, \left(\vf\left(\vm_{k \mid k} + \ML_{k \mid k}\,\vxi_i\right) - \vm_{k+1 \mid k} \right) \\
                      &\times \left( \vf\left(\vm_{k \mid k} + \ML_{k \mid k}\,\vxi_i \right) - \vm_{k+1 \mid k}\right)^\T \Big\} + \MQ, \\    
            \MD_{k+1} &= \sum_i \left\{ w_i\, \ML_{k \mid k}\,\vxi_j\, \left(\vf\left(\vm_{k \mid k} + \ML_{k \mid k}\,\vxi_i \right) - \vm_{k+1 \mid k} \right) \right\}, \\                         
              \MG_{k} &= \MD_{k+1} \, [\MP_{k+1 \mid k}]^{-1}, \\
       \vm_{k \mid T} &= \vm_{k \mid k} + \MG_{k} \, (\vm_{k+1 \mid T} - \vm_{k+1 \mid k}), \\
       \MP_{k \mid T} &= \MP_{k \mid k} + \MG_{k} \, (\MP_{k+1 \mid T} - \MP_{k+1 \mid k}) \, \MG_{k}^\T.
  \end{split}
\end{equation}

To evaluate expectations with respect to the pairwise smoothing distributions (Eq.~\ref{eq:gaussian-pairwise}), the required $2n$-dimensional sigma-points need to be generated separately as they are not used in the smoother pass. The sigma-points used for the pairwise smoothing distributions are of the form 
\begin{equation}                                                                                                                                                                                                                                                                   \begin{pmatrix} \vx^{(i)}_k \\ \vx^{(i)}_{k-1} \end{pmatrix}  = \begin{pmatrix} \vm^{(i)}_{k\mid T} \\ \vm^{(i)}_{k-1\mid T} \end{pmatrix} + \sqrt{\begin{pmatrix}                                                                                                                                                                                                                                                
        \MP_{k \mid T}          & \MP_{k \mid T} \MG_{k-1}^\T \\                                                                                                                                                                                                                         \MG_{k-1}\MP_{k \mid T} & \MP_{k-1 \mid T}                                                                                                                                                                                                                                
      \end{pmatrix}} \, \vxi^{(2n)}_i,                                                             
\label{eq:sigma-point-pairwise}                                                
\end{equation}
where $\vxi^{(2n)}_i$ are the $2n$-dimensional unit sigma-points. Then, expectation a function $\vf(\vx_k,\vx_{k-1})$ is approximated as 
\begin{equation}
\mathbb{E}(\vf(\vx_k,\vx_{k-1}) \mid \vy_{1:T}) = \sum_i w^{(2n)}_i\,\vf(\vx_k^{(i)},\vx_{k-1}^{(i)}), 
\end{equation}
where $w_i^{(2n)}$ are the corresponding weights of the $2n$-dimensional sigma-point scheme.

\section{Parameter Estimation}
\label{sec:parameter-estimation}
In this section, we consider methods for estimating the static parameters $\vtheta$ of the state-space model \eqref{eq:dmodel}. All methods discussed target the maximum likelihood solution, that is, aim to maximize $p(\vy_{1:T} \mid \vtheta)$, or equivalently the log-likelihood:
\begin{equation}
  \vtheta_\mathrm{ML}  = \mathrm{arg\,max}_{\vtheta} \log p(\vy_{1:T} \mid \vtheta).
\end{equation}
Since the state variables $\vx_{0:T}$ cannot in general be marginalized in closed-form, approximative numeric methods are needed. 

Here, we focus on three approaches where sigma-point filtering and smoothing is used to approximate the likelihood. First, we consider a so-called direct-likelihood approach, where the sigma-point algorithm is used to directly approximate the log-likelihood and its gradient, which are then used in numeric optimization algorithms such as conjugate-gradient optimization. Second, the expectation--maximization (EM) algorithm which is based on a lower bound for the log-likelihood and iterating optimization of parameters with respect to the lower bound and updating the lower bound with new parameters. The third approach is a modification of the direct-likelihood optimization where Fisher's identity is used to express the gradient of the log-likelihood using the same lower bound function that appears in the EM algorithm. Note that the third approach is otherwise similar to the first, but since it is based on the EM lower bound, we present the methods in this order. Each of these three approaches may be used in combination with any of the sigma-point rules discussed in Section~\ref{sec:sigmapointschemes}. 

Note that all the algorithms presented in this section are easily extended to maximum \textit{a~posteriori} estimation since maximizing the posterior density is equivalent to maximizing the (unnormalized) log-posterior. That is, the sum of log-likelihood and log-prior:
\begin{equation}
  \vtheta_\mathrm{MAP} = \mathrm{arg\,max}_{\vtheta} \left[
    \log p(\vy_{1:T} \mid \vtheta) + \log p(\vtheta) \right].
\end{equation}
Since the log-prior is known, approximations of the unnormalized log-posterior as well as its gradient and lower bounds are immediately obtained from the corresponding approximations for the log-likelihood.

\subsection{Direct Likelihood Based Parameter Estimation}
\label{sec:direct}
The marginal log-likelihood function can be formulated as the following sum
\begin{equation} \label{eq:energy-function}
  \mathcal{L}_T(\vtheta) = \sum_{k=1}^T \log p(\vy_k \mid \vy_{1:k-1},\vtheta).
\end{equation}
Furthermore, the terms of the sum on the right hand side may in principle be evaluated by  
\begin{equation}
  p(\vy_k \mid \vy_{1:k-1} , \vtheta) = 
    \int p(\vy_k \mid \vx_k, \vtheta) \, 
      p(\vx_{k} \mid \vy_{1:k-1},\vtheta) \dd\vx_k,
\end{equation}
that is, integrating the measurement model $p(\vy_k \mid \vx_k, \vtheta)$ over the predicted state distribution $p(\vx_k \mid \vy_{1:k-1},\vtheta)$ which is computed during the Bayesian filtering recursion. In assumed density Gaussian filtering is used, we get the approximation
\begin{equation}
  p(\vy_k \mid \vy_{1:k-1} , \vtheta) \approx \N(\vy_k \mid \vmu_k, \MS_k),
\end{equation}
whence the marginal log-likelihood expression in Equation~\eqref{eq:energy-function} evaluates to 
\begin{multline}
  \mathcal{L}_T(\vtheta) = 
   \log p(\vy_{1:T} \mid \vtheta) 
    \approx 
    -\frac{1}{2} \sum_{k=1}^T \log |2\pi \, \MS_k| \\
    - \frac{1}{2} \sum_{k=1}^T \left( \vy_k - \vmu_k \right)^\T \MS_k^{-1} 
                               \left( \vy_k - \vmu_k \right),
\end{multline}
where the quantities $\vmu_k$ and $\MS_k$ are evaluated during the filtering recursion, in the case of sigma-point methods by Equation~\eqref{eq:spfupda}.

To enable use of gradient-based optimization algorithms, we also need a method for evaluating the gradients of the marginal log-likelihood. This is based on the so-called sensitivity equations \cite{Goodrich+Caines:1979,Sarkka:2013} that are obtained by differentiating the filtering equations. Namely, the gradient of the log-likelihood is obtained by the recursion
\allowdisplaybreaks
\begin{align}
  \frac{\partial \mathcal{L}_k(\vtheta)}{\partial \theta_i} &= 
    \frac{\partial \mathcal{L}_{k-1}(\vtheta)}{\partial \theta_i} 
    - \frac{1}{2}\tr \left(\MS_k^{-1}(\vtheta) \, \frac{ \partial \MS_k(\vtheta)}{\partial \theta_i} \right) \nonumber \\
    & - \vv_k^\T(\vtheta)\,\MS_k^{-1}(\vtheta)\, \frac{\partial \vv_k(\vtheta)}{\partial \theta_i} \nonumber \\ 
    & + \frac{1}{2} \vv_k^\T(\vtheta) \, \MS_k^{-1}(\vtheta) \, \frac{ \partial \MS_k(\vtheta)}{\partial \theta_i} \, \MS_k^{-1}(\vtheta) \, \vv_k(\theta),
\end{align} 
where $\vv_k = \vy_k - \vmu_k$. The derivatives $\frac{\partial \MS_k(\vtheta)}{\partial \theta_i}$ and $\frac{\partial \vv_k(\vtheta)}{\partial \theta_i}$ are computed along the filtering pass by the equations shown in Figure~\ref{fig:sensitivityequations}.
\begin{figure}
\allowdisplaybreaks
\footnotesize
\begin{align*}
&\frac{\partial \vm_{k\mid k-1}}{\partial \theta_i} = 
  \sum_j \bigg\{ w_j \,  \bigg[ \MF_\vx \left( \vm_{k-1\mid k-1} + \ML_{k-1 \mid k-1}\,\vxi_j , ~ \vtheta \right)  \nonumber \\
  &\qquad \times \bigg( \frac{\partial \vm_{k-1 \mid k-1}}{\partial \theta_i} + \frac{\partial \ML_{k-1 \mid k-1}}{\partial \theta_i}\,\vxi_j \bigg) \nonumber \\
  &\qquad +  \frac{\partial \vf}{\partial \theta_i}\left(\vm_{k-1 \mid k-1} + \ML_{k-1}\,\vxi_j,~ \vtheta \right) \bigg] \bigg\}, \\
&\frac{\partial \MP_{k \mid k-1}}{\partial \theta_i} = 
  \sum_j \bigg\{ w_j \, \bigg[ \MF_\vx \left( \vm_{k-1} + \ML_{k-1}\, \vxi_j, \vtheta \right)  \nonumber \\
  &\qquad \times  \bigg( \frac{\partial \vm_{k-1 \mid k-1}}{\partial \theta_i} + \frac{\partial \ML_{k-1 \mid k-1}}{\partial \theta_i}\,\vxi_j \bigg) \nonumber \\
  &\qquad +  \frac{\partial \vf}{\partial \theta_i} \left(\vm_{k-1 \mid k-1} + \ML_{k-1\mid k-1} \, \vxi_j, \vtheta \right) - \frac{\partial \vm_{k \mid k-1}}{\partial \theta_i} \bigg] \nonumber \\
  &\qquad \times  \left[ \vf \left( \vm_{k-1 \mid k-1} + \ML_{k-1 \mid k-1} \, \vxi_j , \vtheta \right) - \vm_{k \mid k-1} \right]^\T \nonumber \\
  &\qquad + \left[ \vf \left( \vm_{k-1 \mid k-1} + \ML_{k-1 \mid k-1} \, \vxi_j , \vtheta \right) - \vm_{k \mid k-1} \right] \nonumber \\
  &\qquad \times \bigg[ \MF_\vx \left( \vm_{k-1} + \ML_{k-1}\, \vxi_j, \vtheta \right)  \nonumber \\
  &\qquad \times \bigg( \frac{\partial \vm_{k-1 \mid k-1}}{\partial \theta_i} + \frac{\partial \ML_{k-1 \mid k-1}}{\partial \theta_i}\,\vxi_j \bigg) \nonumber \\
  &\qquad +  \frac{\partial \vf}{\partial \theta_i} \left(\vm_{k-1 \mid k-1} + \ML_{k-1\mid k-1} \, \vxi_j, \vtheta \right) - \frac{\partial \vm_{k \mid k-1}}{\partial \theta_i} \bigg]^\T  +  \frac{\partial \MQ}{\partial \theta_i} \bigg\}, \\
&\frac{\partial \vmu_k}{\partial \theta_i} = 
  \sum_j \bigg\{ w_j \bigg[ \MH_\vx \left( \vm_{k \mid k-1} + \ML_{k \mid k-1} \, \vxi_j,~\vtheta \right)   \nonumber \\
  &\qquad + \frac{\partial \vh}{\partial \theta_i}\left( \vm_{k \mid k-1} + \ML_{k \mid k-1}\, \vxi_j,~\vtheta \right) \bigg] \bigg\}, \\
& \frac{\partial \vv_k}{\partial \theta_i} = 
  -\frac{\partial \vmu_k}{\partial \theta_i}, \\
&\frac{\partial \MS_k}{\partial \theta_i} = 
  \sum_j \bigg\{ w_j \, \bigg[ \MH_\vx \left( \vm_{k \mid k-1} + \ML_{k \mid k-1}\, \vxi_j,~\vtheta\right)  \nonumber \\
  &\qquad \times \bigg(\frac{\partial \vm_{k \mid k-1}}{\partial \theta_i} + \frac{\partial \ML_{k \mid k-1}}{\partial \theta_i} \bigg) \nonumber \\
  &\qquad +  \frac{\partial \vh}{\partial \theta_i} \left(\vm_{k \mid k-1} + \ML_{k \mid k-1}\,\vxi_j,~\vtheta \right) - \frac{\partial \vmu_k}{\partial \theta_i} \bigg] \nonumber \\
  &\qquad \times \bigg[ \vh \left( \vm_{k \mid k-1} + \ML_{k \mid k-1} \, \vxi_j,~ \vtheta \right) - \vmu_k \bigg]^\T \nonumber \\
  &\qquad + \bigg[ \vh \left( \vm_{k \mid k-1} + \ML_{k \mid k-1} \, \vxi_j,~ \vtheta \right) - \vmu_k \bigg] \nonumber \\
  &\qquad \times \bigg[ \MH_\vx \left( \vm_{k \mid k-1} + \ML_{k \mid k-1}\, \vxi_j,~\vtheta\right)  \nonumber \\
  &\qquad +   \frac{\partial \vh}{\partial \theta_i} \left(\vm_{k \mid k-1} + \ML_{k \mid k-1}\,\vxi_j \right) - \frac{\partial \vmu_k}{\partial \theta_i} \bigg]^\T \bigg\} + \frac{\partial \MR}{\partial \theta_i}, \\
&\frac{\partial \MC_k}{\partial \theta_i} = 
  \sum_j \bigg\{ w_j  \, \bigg[ \frac{\partial \ML_{k \mid k-1}}{\partial \theta_i}\,\vxi_j\,\bigg(\vh\left( \vm_{k \mid k-1} + \ML_{k \mid k-1}\,\vxi_j, \vtheta\right) - \mu_k \bigg)^\T  \nonumber \\
  &\qquad + \ML_{k \mid k-1}\,\vxi_j\,\bigg[ \MH_\vx\, \left(\vm_{k \mid k-1} + \ML_{k \mid k-1}\,\vxi_j,~\vtheta\right) \nonumber \\
  &\qquad \times \bigg(\frac{\partial \vm_{k \mid k-1}}{\partial \theta_i} + \frac{\partial \ML_{k \mid k-1}}{\partial \theta_i}\,\vxi_j \bigg) \nonumber \\
  &\qquad +   \frac{\partial \vh}{\partial \theta_i}\left(\vm_{k \mid k-1} + \ML_{k \mid k-1}\,\vxi_j,~\vtheta \right) - \frac{\partial \vmu_k}{\partial \theta_i}\bigg]^\T \bigg] \bigg\}, \\
&\frac{\partial \MK_k}{\partial \theta_i} = 
  \frac{\partial \MC_k}{\partial \theta_i}\,\MS_k^{-1} - \MC_k\,\MS_k^{-1}\,\frac{\partial \MS_k}{\partial \theta_i}\,\MS_k^{-1}, \\
&\frac{\partial \vm_{k \mid k}}{\partial \theta_i} = 
  \frac{\partial \vm_{k \mid k-1}}{\partial \theta_i} + \frac{\partial \MK_k}{\partial \theta_i}\,\vv_k + \MK_k\,\frac{\partial \vv_k}{\partial \theta_i}, \\
&\frac{\partial \MP_{k \mid k}}{\partial \theta_i} = 
  \frac{\partial \MP_{k \mid k-1}}{\partial \theta_i} - \frac{\partial \MK_k}{\partial \theta_i}\,\MS_k\,\MK_k^\T - \MK_k\,\frac{\partial \MS_k}{\partial \theta_i}\,\MK_k^\T -\MK_k\,\MS_k\,\frac{\partial \MK_k^\T}{\partial \theta_i}.
\end{align*}
\caption{Recursion for computing the derivatives of the prediction and update steps. $\MF_\vx$ is the Jacobian of $\vf(\vx,\vtheta)$ as a function of $\vx$ and $\MH_\vx$ is the Jacobian of $\vh(\vx,\vtheta)$ as a function of $\vx$. Algorithms for computing the derivatives of the Cholesky factors $\ML$ such that $\MP = \ML\,\ML^\T$ are omitted here (see \cite{Sarkka:2013}).}
\label{fig:sensitivityequations}
\end{figure}

\subsection{Expectation--Maximization Based Parameter Estimation}
\label{sec:em}

Expectation--maximization (EM), proposed by Dempster \etal\ \cite{Dempster:1977}  is an iterative algorithm for finding maximum likelihood parameter estimates in settings with some unobserved variables, such as the state variables $\vx$ in the state-space context. The motivation is that the so-called full-data likelihood of the observed and unobserved variables is easier to compute, and a lower bound for the marginal likelihood of the observed variables may be obtained based on expected full-data log-likelihood. In the following, we present the EM algorithm following the formulation by Neal and Hinton \cite{Neal+Hinton:1998} and the notation of Sch\"on \etal\ \cite{Schon+Wills+Ninness:2011}. 

The EM algorithm is based on the following lower bound of the log-likelihood:
\begin{equation} \label{eq:emlowerbound}
  \log p(\vy_{1:T} \mid \vtheta) \geq 
    \int q(\vx_{0:T}) \log \frac{p(\vx_{0:T}, \vy_{1:T} \mid \vtheta)}{q(\vx_{0:T})} \dd \vx_{0:T},
\end{equation}
where $q$ is an arbitrary probability density over the states $\vx_{0:T}$. The idea is to iteratively maximize this lower bound with respect to $q$ (holding $\vtheta$ fixed) and with respect to $\vtheta$ (holding $q$ fixed). Furthermore, when $\vtheta = \vtheta^{(n)}$ is fixed, the maximum with respect to $q$ is obtained by 
\begin{equation}
 q(\vx_{0:T}) := p(\vx_{0:T} \mid \vy_{1:T}, \vtheta^{(n)}).
\end{equation}
By substituting this into Equation~\eqref{eq:emlowerbound}, the bound becomes
\begin{align*}
  & \int p(\vx_{0:T} \mid \vy_{1:T}, \vtheta^{(n)}) \log \frac{p(\vx_{0:T}, \vy_{1:T} \mid \vtheta)}{p(\vx_{0:T} \mid \vy_{1:T}, \vtheta^{(n)})} \dd \vx_{0:T} \nonumber \\ 
  & \quad = \int p(\vx_{0:T} \mid \vy_{1:T}, \vtheta^{(n)}) \log p(\vx_{0:T},\vy_{1:T} \mid \vtheta) \dd \vx_{0:T} \nonumber \\
  & \qquad  - \int p(\vx_{0:T} \mid \vy_{1:T}, \vtheta^{(n)}) \log p(\vx_{0:T} \mid \vy_{1:T}, \vtheta^{(n)}) \dd \vx_{0:T}.
\end{align*}

The latter term is independent of $\vtheta$ and may thus be omitted when maximizing the lower bound with respect to $\vtheta$. The first term is the conditional expectation of $\log p(\vy_{1:T},\vx_{0:T} \mid \vtheta)$ conditional on $\vtheta^{(n)}$ and $\vy_{1:T}$. Thus, the step of maximizing the lower bound (Eq.~\ref{eq:emlowerbound}) may be replaced by computing the following function:
\begin{equation}
  \mathcal{Q}(\vtheta,\vtheta^{(n)}) = \E[\log p(\vx_{0:T}, \vy_{1:T} \mid \vtheta) \mid \vy_{1:T},\vtheta^{(n)}].
\end{equation}
The EM algorithm in its general form thus consists of initializing the parameters to $\vtheta^{(0)}$ and for $n=0,1,\ldots$ iterating the following two steps:
\begin{itemize}
 \item E-step: compute $\mathcal{Q}(\vtheta,\vtheta^{(n)})$.
 \item M-step: $\vtheta^{(n+1)} \gets \argmax_{\vtheta} \mathcal{Q}(\vtheta,\vtheta^{(n)})$.
\end{itemize}
In state-space models, the $\mathcal{Q}$-function can be decomposed by employing the Markov property of the state sequence and the conditional independence of the measurements:
\begin{equation}
\label{eq:Q_decomposition}
  \mathcal{Q}(\vtheta, \vtheta^{(n)}) = \mathrm{I}_1(\vtheta, \vtheta^{(n)}) + \mathrm{I}_2(\vtheta, \vtheta^{(n)}) + \mathrm{I}_3(\vtheta, \vtheta^{(n)}),
\end{equation}
where the terms are 
\begin{align} \label{eq:Q_I1}
  \mathrm{I}_1(\vtheta, \vtheta^{(n)}) &=
     \E[\log p(\vx_0 \mid \vtheta ) \mid \vy_{1:T},\vtheta^{(n)}], \vphantom{\sum_{k}^{T}} \\ \label{eq:Q_I2}
  \mathrm{I}_2(\vtheta, \vtheta^{(n)}) &=
     \sum_{k=1}^{T}\E[\log p(\vx_{k} \mid \vx_{k-1}, \vtheta) \mid \vy_{1:T}, \vtheta^{(n)}], \\ \label{eq:Q_I3}
  \mathrm{I}_3(\vtheta, \vtheta^{(n)}) &=
     \sum_{k=1}^T \E[\log p(\vy_k \mid \vx_k, \vtheta) \mid \vy_{1:T}, \vtheta^{(n)}].
\end{align}
To evaluate this expression, one needs the smoothing distributions $p(\vx_t \mid \vy_{1:T}, \vtheta^{(n)})$ and the joint smoothing distributions of consecutive states $p(\vx_k, \vx_{k+1} \mid \vy_{1:T},\vtheta^{(n)})$. Sigma-point approximations to the EM algorithm are then obtained by replacing the expectations over the smoothing distributions by their sigma-point smoother approximations. The Gaussian smoother approximation for $\mathcal{Q}$ is 
\begin{align}
  &\mathcal{Q}(\vtheta,\vtheta^{(n)}) \nonumber \\ 
  &\approx -\frac{1}{2} \log |2\pi\,\MP_0|
  - \frac{T}{2} \log |2\pi\,\MQ| 
  - \frac{T}{2} \log |2\pi\,\MR| \nonumber \\ 
  &  -\frac{1}{2} \tr \Big\{ \MP_0^{-1} \,  \Big[ \MP_{0 \mid T} + (\vm_{0 \mid T} - \vm_0) \, (\vm_{0 \mid T} - \vm_0)^\T
      \Big] \Big\} \nonumber \\ 
  &  -\frac{1}{2} \sum_{k=1}^T
     \tr \big\{ \MQ^{-1} \,  \E\big[ (\vx_{k} - \vf(\vx_{k-1})) 
                   (\vx_{k} - \vf(\vx_{k-1}))^\T
       \mid  \vy_{1:T} \big] \big\} \nonumber \\
  &  -\frac{1}{2} \sum_{k=1}^T
     \tr \big\{ \MR^{-1} \,   \E\big[ (\vy_{k} - \vh(\vx_{k})) 
                   (\vy_{k} - \vh(\vx_{k}))^\T
       \mid  \vy_{1:T} \big] \big\},
\label{eq:gaussiansmootherQ}       
\end{align}
where $\MP_0,\MQ,\MR$ and the model functions $\vf(\cdot),\vh(\cdot)$ depend on the parameters $\vtheta$. The smoothing distribution means and covariances $\vm_{k \mid T},\MP_{k \mid T}$ are obtained during the smoothing backward pass. The expectations over the smoothing distribution in the latter two terms are evaluated by using the sigma-point approximations for Gaussian integrals as follows. The second expectation depends only on the smoothing distribution $\N(\vx_k \mid \vm_{k \mid T}, \MP_{k \mid T})$ and is computed as follows:
\begin{multline}
  \E\big[ (\vy_{k} - \vh(\vx_{k}) (\vy_{k} - \vh(\vx_{k})^\T
    \mid  \vy_{1:T} \big] \vphantom{\sum_i} \\
    \approx \sum_i w_i\, (\vy_{k} - \vh(\vm_{k \mid T}) + \ML_{k \mid T}\vxi_i) \\
    \times (\vy_{k} - \vh(\vm_{k \mid T}) + \ML_{k \mid T}\vxi_i)^\T.
\end{multline}
The first expectation depends on the pairwise joint smoothing distribution $p(\vx_k, \vx_{k-1} \mid \vy_{1:T})$ (\cf\ Sec.~\ref{sec:gaussianfiltering}, Eq.~\ref{eq:gaussian-pairwise}). Thus, to evaluate it we need to use $2n$-dimensional sigma-points as discussed in Section~\ref{sec:sigmapointfiltering}, Equation~\ref{eq:sigma-point-pairwise}.

In general, maximizing $\mathcal{Q}$ in the M-step requires the use numerical optimization, for example, using the Broyden--Fletcher--Goldfarb--Shanno (BFGS) algorithm \cite{Fletcher:1987}. However, using numerical optimization inside EM is quite cumbersome, because with the same effort we could numerically optimize the approximate likelihood directly. Hence the benefit of EM is in the situation when the optimization can be performed in closed form. This kind of special case is the class of models where the parameters appear linearly although the model itself might be nonlinear.

In the following, we present closed-form solutions for the special case where the model functions are linear combinations of the parameters where the parameters appear as coefficients of the linear combinations and/or the covariances. That is, we consider models that can be represented as follows:
\begin{align}
  \vx_k &=  \MA \, \tilde{\vf}(\vx_{k-1}) + \vq_k, \\
  \vy_k &= \MH \, \tilde{\vh}(\vx_k) + \vr_k,
\end{align}
where $\tilde{\vf}(\cdot)$ and $\tilde{\vh}(\cdot)$ are functions containing the nonlinearities and the parameters are a subset of $\{\MA,\MH,\MQ,\MR,\vm_0,\MP_0\}$.

For these models, the expression for $\mathcal{Q}$ can be written as 
\begin{align}
  &\mathcal{Q}(\vtheta,\vtheta^{(n)}) = \nonumber \\
  & - \frac{1}{2} \log |2\pi\,\MP_0|
    - \frac{T}{2} \log |2\pi\,\MQ| 
    - \frac{T}{2} \log |2\pi\,\MR| \nonumber \\
  & - \frac{1}{2} \tr \Big\{ \MP_0^{-1} \Big[
      \MP_{0 \mid T} + (\vm_{0 \mid T} - \vm_0) \, (\vm_{0 \mid T} - \vm_0)^\T
      \Big] \Big\} \nonumber \\
  & - \frac{T}{2}
     \tr \Big\{ \MQ^{-1}  \Big[ \MSigma
     -  \MC \, \MA^\T 
     - \MA \, \MC^\T 
     + \MA \, \MPhi
      \, \MA^\T
      \Big] \Big\} \nonumber \\
  &  -\frac{T}{2} 
     \tr \Big\{ \MR^{-1} \nonumber \Big[
      \MD
    -  \MB \, \MH^\T
    - \MH \, \MB^\T
    + \MH \, \MTheta \, \MH^\T
      \Big] \Big\},
\end{align}
where the model parameters to be optimized are some subset of $\{\MA,\MH,\MQ,\MR\,\vm_0,\MP_0\}$ and $\MSigma,\MPhi,\MTheta,\MB,\MC,\MD$ can be evaluated based on the latest E-step sigma-point smoother results as follows:
\begin{align}
 \MSigma  &= \frac{1}{T} \sum_{k=1}^T  \label{eq:MSigma}
             \MP_{k \mid T} + \vm_{k \mid T} \, [\vm_{k \mid T}]^\T, \\
    \MPhi &= \frac{1}{T} \sum_{k=1}^T 
             \E\big[ \tilde{\vf}(\vx_{k-1}) \, 
                     \tilde{\vf}^\T(\vx_{k-1}) \mid \vy_{1:T} \big], \\
  \MTheta &= \frac{1}{T} \sum_{k=1}^T 
             \E\big[ \tilde{\vh}(\vx_{k}) \, 
                     \tilde{\vh}^\T(\vx_{k}) \mid \vy_{1:T} \big], \\
    \MB   &= \frac{1}{T} \sum_{k=1}^T 
             \vy_k \, \E\big[ \tilde{\vh}^\T(\vx_{k}) \mid \vy_{1:T} \big], \\
    \MC   &= \frac{1}{T} \sum_{k=1}^T 
             \E\big[ \vx_{k} \, \tilde{\vf}^\T(\vx_{k-1}) \mid \vy_{1:T} \big], \\
    \MD   &= \frac{1}{T} \sum_{k=1}^T \vy_k \, \vy_k^\T. \label{eq:MD}
\end{align}
Using these values, the optimal parameters in the M-step, that is, the maximum points of the $\mathcal{Q}(\cdot,\vtheta^{(n)})$-function are 
\begin{itemize}
\item When $\vtheta = \MA$, we get
\begin{equation*} \label{eq:linqmax_a}
  \MA^* = \MC \, \MPhi^{-1}.
\end{equation*}
\item When $\vtheta = \MH$, we get
\begin{equation*} \label{eq:linqmax_h}
  \MH^* = \MB \, \MTheta^{-1}.
\end{equation*}
\item When $\vtheta = \MQ$, we get
\begin{equation*} \label{eq:linqmax_q} 
  \MQ^* = \MSigma - \MC \, \MA^\T - \MA \, \MC^\T + \MA \, \MPhi \, \MA^\T.
\end{equation*}
\item When $\vtheta = \MR$, we get
\begin{equation*} \label{eq:linqmax_r}
  \MR^* = \MD - \MH \, \MB^\T - \MB \, \MH^\T + \MH \, \MTheta \, \MH^\T.
\end{equation*}
\item When $\vtheta = \vm_0$, we get
\begin{equation} \label{eq:linqmax_m0}
  \vm_0^* = \vm_{0 \mid T}.
\end{equation}
\item Finally, the maximum with respect to the initial covariance $\vtheta = \MP_0$ is
\begin{equation*} \label{eq:linqmax_p0}
  \MP_0^* = \MP_{0 \mid T} + (\vm_{0 \mid T} - \vm_0) \, (\vm_{0 \mid T} - \vm_0)^\T.
\end{equation*}
\end{itemize}

\subsection{Evaluating the Gradient Based on Fisher's Identity}
\label{sec:fisher}
The expected log-likelihood that appears in the EM algorithm may also be used as a basis of an alternative approach for evaluating the gradient in direct optimization. Based on Fisher's identity, the gradient of the marginal log-likelihood may be expressed as 
\begin{equation}
  \frac{\partial \mathcal{L}_T(\vtheta)}{\partial \vtheta} = \left.\frac{\partial \mathcal{Q}(\vtheta,\vtheta^{(n)})}{\partial \vtheta } \right|_{\vtheta^{(n)}=\vtheta},
\end{equation}
where $\mathcal{Q}$ is the function defined in the EM algorithm (Eq.~\ref{eq:Q_decomposition}). When the $\mathcal{Q}$-function is approximated with sigma-point smoothers, we obtain an alternative approximation of the gradient of the marginal log-likelihood that may be used in place of the approximation derived in Section~\ref{sec:direct}. For linear state-space models, this approach was suggested by Segal and Weinstein \cite{Segal+Weinstein:1988} and later by Olsson \etal\ \cite{Olsson+Petersen+Lehn-Schioler:2007} who called the approach the `easy gradient recipe'. See \cite{Cappe+Moulines:2005,Sarkka:2013} for discussions of the nonlinear case.

\section{Experiments}
In this section, we demonstrate the different sigma-point schemes and different parameter estimation algorithms with two example models. First, we use a one-dimensional model (the univariate nonstationary growth model, UNGM, \cite{Andrade-Netto+Gimeno+Mendes:1979,Kitagawa:1987-rejoinder}) to illustrate the approximate likelihood curves obtained by different methods. Second, we compare the performance of different algorithms with simulated data in a problem of tracking a maneuvering target with bearings-only measurements. In this example, we focus on estimating the sensor variances and compare the variance estimates as well as the actual tracking error. 

\subsection{Simple Nonlinear Growth Model}
We simulated a realization with $T=100$ data from the following model:
\begin{align}
  x_{k+1} &=  a\,x_k + b\,\frac{x_k}{1+x_k^2} + c\,\cos(1.2k) + q_k, \\
   y_k &= d\,x_k + r_k.
\end{align}
with $a=0.5$, $b=25$, $c=8$, $d=\sqrt{0.05}$, $q_k \sim \N(0,10)$, $r_k \sim \N(0,0.01)$, $x_0 \sim \N(0,0.01)$. This is the univariate nonstationary growth model \cite{Andrade-Netto+Gimeno+Mendes:1979, Kitagawa:1987-rejoinder} except that we changed the measurement model to linear as the model with the typically used quadratic measurement model is known to be challenging for sigma-point algorithms \cite{Wu+Hu+Wu+Hu:2005}. 

\begin{figure}[!t]
  \pgfplotsset{
    compat=newest,    
    hide y axis,
    tick align=outside,
    minor tick num=1,
    legend style={nodes=right,text width=6em,row sep=1em, font=\scriptsize, xshift=1em},
  }
  %
  %\tikzsetnextfilename{fig04}
  \setlength{\figurewidth}{0.9\columnwidth}
  \setlength{\figureheight}{0.6\columnwidth}
  \centering \sffamily \scriptsize
  % This file was created by matlab2tikz v0.4.2.
% Copyright (c) 2008--2013, Nico Schlömer <nico.schloemer@gmail.com>
% All rights reserved.
% 
% The latest updates can be retrieved from
%   http://www.mathworks.com/matlabcentral/fileexchange/22022-matlab2tikz
% where you can also make suggestions and rate matlab2tikz.
% 
% 
% 
\begin{tikzpicture}

\begin{axis}[%
width=\figurewidth,
height=\figureheight,
unbounded coords=jump,
scale only axis,
xmin=-15,
xmax=15,
xlabel={Parameter $a$},
ymin=-24156.9824437913,
ymax=1492.66755843199,
ytick={\empty},
axis x line*=bottom,
axis y line*=left,
legend style={draw=black,fill=white,legend cell align=left}
]
\addplot [
color=black,
dotted,
line width=1.2pt,
line cap=round, dash pattern=on 0pt off 2\pgflinewidth
]
table[row sep=crcr]{
-15 -inf\\
-14.0322580645161 -inf\\
-13.0645161290323 -inf\\
-12.0967741935484 -inf\\
-11.1290322580645 -inf\\
-10.1612903225806 -inf\\
-9.19354838709677 -inf\\
-8.2258064516129 -inf\\
-7.25806451612903 -inf\\
-6.29032258064516 -inf\\
-5.32258064516129 -inf\\
-4.35483870967742 -9956.83402492046\\
-3.38709677419355 -6628.05734336311\\
-2.41935483870968 -3899.48128835694\\
-1.45161290322581 -1874.14343390616\\
-0.483870967741936 -546.123216283621\\
0.483870967741936 -75.7372944517991\\
1.45161290322581 -590.221889119951\\
2.41935483870968 -2049.88370559498\\
3.38709677419355 -4459.88011989549\\
4.35483870967742 -7990.99851462041\\
5.32258064516129 -12595.9251603017\\
6.29032258064516 -inf\\
7.25806451612903 -inf\\
8.2258064516129 -inf\\
9.19354838709678 -inf\\
10.1612903225806 -inf\\
11.1290322580645 -inf\\
12.0967741935484 -inf\\
13.0645161290323 -inf\\
14.0322580645161 -inf\\
15 -inf\\
};
\addlegendentry{Likelihood (particle)};

\addplot [
color=blue,
solid,
line width=1.2pt
]
table[row sep=crcr]{
-15 -21812.9961108364\\
-14.0322580645161 -21268.8042127425\\
-13.0645161290323 -20643.585310257\\
-12.0967741935484 -19922.3205478077\\
-11.1290322580645 -19087.2238822367\\
-10.1612903225806 -18117.599822142\\
-9.19354838709677 -16990.1065595392\\
-8.2258064516129 -15679.8593949888\\
-7.25806451612903 -14163.1197441093\\
-6.29032258064516 -12422.7431482329\\
-5.32258064516129 -10457.9861506891\\
-4.35483870967742 -8300.25162336228\\
-3.38709677419355 -6034.72329804911\\
-2.41935483870968 -3822.34661941984\\
-1.45161290322581 -1906.39267045001\\
-0.483870967741936 -580.989931341186\\
0.483870967741936 -110.435566706972\\
1.45161290322581 -624.302998477215\\
2.41935483870968 -2053.04686514606\\
3.38709677419355 -4155.99055401445\\
4.35483870967742 -6619.88377694834\\
5.32258064516129 -9160.76870055458\\
6.29032258064516 -11579.0422252286\\
7.25806451612903 -13764.7960971829\\
8.2258064516129 -15676.8682581806\\
9.19354838709678 -17316.7538585128\\
10.1612903225806 -18707.7581936316\\
11.1290322580645 -19881.567929973\\
12.0967741935484 -20870.8184364276\\
13.0645161290323 -21705.5547162906\\
14.0322580645161 -22411.9084126288\\
15 -23011.9087829778\\
};
\addlegendentry{Likelihood (sigma-point)};

\addplot [
color=red,
dashed,
line width=1.0pt
]
table[row sep=crcr]{
-15 -21729.9954787635\\
-14.0322580645161 -20831.7091136738\\
-13.0645161290323 -20002.7221384243\\
-12.0967741935484 -19243.0345530151\\
-11.1290322580645 -18552.6463574461\\
-10.1612903225806 -17931.5575517174\\
-9.19354838709677 -17379.7681358289\\
-8.2258064516129 -16897.2781097807\\
-7.25806451612903 -16484.0874735727\\
-6.29032258064516 -16140.1962272049\\
-5.32258064516129 -15865.6043706774\\
-4.35483870967742 -15660.3119039901\\
-3.38709677419355 -15524.3188271431\\
-2.41935483870968 -15457.6251401363\\
-1.45161290322581 -15460.2308429698\\
-0.483870967741936 -15532.1359356435\\
0.483870967741936 -15673.3404181575\\
1.45161290322581 -15883.8442905117\\
2.41935483870968 -16163.6475527062\\
3.38709677419355 -16512.7502047409\\
4.35483870967742 -16931.1522466158\\
5.32258064516129 -17418.853678331\\
6.29032258064516 -17975.8544998864\\
7.25806451612903 -18602.1547112821\\
8.2258064516129 -19297.754312518\\
9.19354838709678 -20062.6533035942\\
10.1612903225806 -20896.8516845106\\
11.1290322580645 -21800.3494552673\\
12.0967741935484 -22773.1466158642\\
13.0645161290323 -23815.2431663014\\
14.0322580645161 -24926.6391065787\\
15 -26107.3344366964\\
};
\addlegendentry{EM bound};

\addplot [
color=red,
dashed,
line width=1.0pt,
forget plot
]
table[row sep=crcr]{
-15 -101293.904398004\\
-14.0322580645161 -89014.5612800962\\
-13.0645161290323 -77532.626860812\\
-12.0967741935484 -66848.1011401507\\
-11.1290322580645 -56960.9841181129\\
-10.1612903225806 -47871.2757946978\\
-9.19354838709677 -39578.9761699061\\
-8.2258064516129 -32084.0852437374\\
-7.25806451612903 -25386.6030161918\\
-6.29032258064516 -19486.5294872694\\
-5.32258064516129 -14383.8646569701\\
-4.35483870967742 -10078.6085252938\\
-3.38709677419355 -6570.76109224076\\
-2.41935483870968 -3860.32235781082\\
-1.45161290322581 -1947.29232200398\\
-0.483870967741936 -831.670984820278\\
0.483870967741936 -513.458346259688\\
1.45161290322581 -992.654406322221\\
2.41935483870968 -2269.25916500787\\
3.38709677419355 -4343.27262231664\\
4.35483870967742 -7214.69477824852\\
5.32258064516129 -10883.5256328036\\
6.29032258064516 -15349.7651859817\\
7.25806451612903 -20613.4134377829\\
8.2258064516129 -26674.4703882073\\
9.19354838709678 -33532.9360372548\\
10.1612903225806 -41188.8103849255\\
11.1290322580645 -49642.0934312192\\
12.0967741935484 -58892.7851761359\\
13.0645161290323 -68940.885619676\\
14.0322580645161 -79786.3947618391\\
15 -91429.3126026252\\
};
\addplot [
color=white!70!black,
solid,
forget plot
]
table[row sep=crcr]{
-12 -24156.9824437913\\
-12 -110.435566706972\\
};
\node[inner sep=0mm, text=black]
at (axis cs:-12,805.623361943861) {$\theta^{(n)}$};
\addplot [
color=white!70!black,
solid,
forget plot
]
table[row sep=crcr]{
-1.97187160698064 -24156.9824437913\\
-1.97187160698064 -110.435566706972\\
};
\node[inner sep=0mm, text=black]
at (axis cs:-1.97187160698064,805.623361943861) {$\theta^{(n+1)}$};
\addplot [
color=gray!60!black,
dotted,
line width=1.2pt,
forget plot,
line cap=round, dash pattern=on 0pt off 2\pgflinewidth
]
table[row sep=crcr]{
-15 -45952.8822255149\\
-14.0322580645161 -45764.61298262\\
-13.0645161290323 -44382.9919921735\\
-12.0967741935484 -42020.8488813468\\
-11.1290322580645 -38880.6764141135\\
-10.1612903225806 -35154.6304912498\\
-9.19354838709677 -31024.5301503342\\
-8.2258064516129 -26661.8575657475\\
-7.25806451612903 -22227.758048673\\
-6.29032258064516 -17873.0400470963\\
-5.32258064516129 -13738.1751458055\\
-4.35483870967742 -9953.29806639097\\
-4.35483870967742 -9956.83402492046\\
-3.38709677419355 NaN\\
-2.41935483870968 NaN\\
-1.45161290322581 NaN\\
-0.483870967741936 NaN\\
0.483870967741936 NaN\\
1.45161290322581 NaN\\
2.41935483870968 NaN\\
3.38709677419355 NaN\\
4.35483870967742 NaN\\
5.32258064516129 -12592.6390390409\\
5.32258064516129 -12595.9251603017\\
6.29032258064516 -18290.176750716\\
7.25806451612903 -25093.0436139387\\
8.2258064516129 -32996.9951287304\\
9.19354838709678 -41987.4499319149\\
10.1612903225806 -52039.4897971184\\
11.1290322580645 -63117.8596347697\\
12.0967741935484 -75176.9674920996\\
13.0645161290323 -88160.8845531416\\
14.0322580645161 -102003.345138731\\
15 -116627.746706507\\
};
\end{axis}
\end{tikzpicture}%
  \caption{Visualization of the one-step evolution of the EM algorithm for
    the univariate estimation of parameter $a$. The dotted line represents the
    particle filter log-likelihood estimate, while the solid line is the sigma-point
    filter log-likelihood approximation. The dashed lines correspond to the sigma-point EM bounds
    for iterations $n$ and $n+1$.}
  \label{fig:sigma-point-em}
\end{figure}
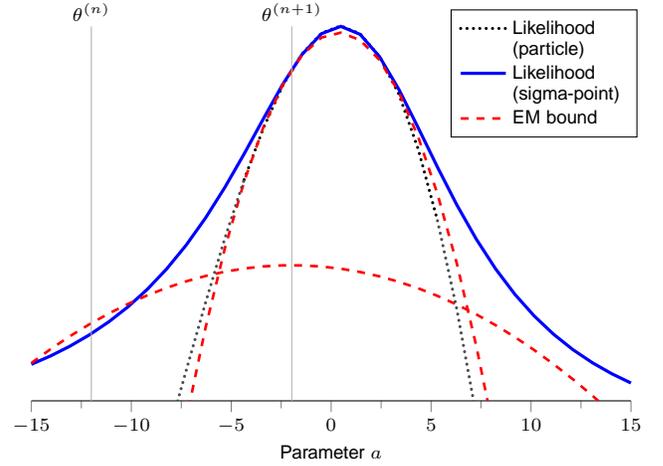

First, we estimated the likelihood of parameter $a$, holding other parameters fixed at their ground-truth values.  Likelihood curves obtained by direct likelihood estimation with various sigma-point rules as well as the EM lower bounds for two iterations are shown in Figure~\ref{fig:sigma-point-em}. For comparison, a likelihood estimate obtained by particle filtering ($1000$ particles and the optimal importance distribution) is also shown. The EM iterations seem to converge toward the maximum of the likelihood curve and the second EM bound is rather close to the particle filter likelihood estimate. The EM lower bounds are mostly below the sigma-point likelihood curve as expected, except that the first EM lower bound slightly exceeds the sigma-point likelihood approximation in the vicinity of the initial parameter. However, both the evaluation of $\mathcal{Q}$ and the sigma-point estimate of the likelihood are approximations. 

\begin{figure}[!t]
  \pgfplotsset{
    compat=newest,    
    clip=false,
    hide y axis,
    tick align=outside,
    minor tick num=1,
    legend style={font=\scriptsize, xshift=-0.25\columnwidth},
  }
  %
  %\tikzsetnextfilename{fig05}
  \setlength{\figurewidth}{0.45\columnwidth}
  \setlength{\figureheight}{0.6\columnwidth}
  \centering \sffamily \scriptsize
  \input{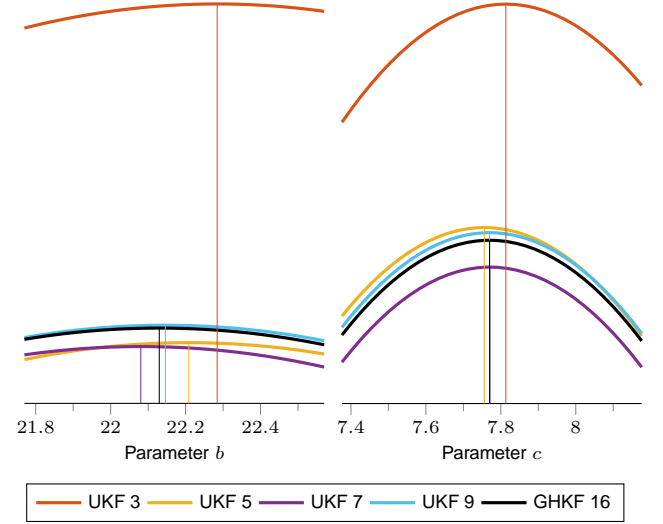}
  \caption{Log-likelihood curves for parameters $b$ and $c$ evaluated by five different sigma-point methods. The vertical line indicates the location of the maximum.}
  \label{fig:sigma-point-lik}
\end{figure}

Second, to compare the different sigma-point rules, we considered estimation of the parameter $b$ with other parameters fixed and parameter $c$ with other parameters fixed, using a grid of parameter values with close proximity to the maximum likelihood values. Namely, for $b$ we used $32$ evenly spaced points between $21.7698$ and $22.5698$ and for $c$ we used $32$ points between $7.376$ and $8.176$. These are shown in Figure~\ref{fig:sigma-point-lik}. The estimate obtained by the Gauss--Hermite rule and the estimates obtained by the higher-order UKFs are rather close to each other while the estimate by the 3$^\text{rd}$ order UKF is farther in both parameters.

\subsection{Coordinated-Turn Model}
In this section, we compare the performance of the parameter estimation methods discussed in this article using a more practical example. The problem is tracking a target maneuvering according to the coordinated turn model \cite{Bar-Shalom+Li+Kirubarajan:2001,Arasaratnam+Haykin:2009,Sarkka+Hartikainen:2010,Solin:2010} with bearings-only sensor measurements. The state is $5$-dimensional:
\begin{equation}
 \vx = \begin{pmatrix}x_1 & x_2 & \dot{x}_1 & \dot{x}_2 & \omega \end{pmatrix}^\T,
\end{equation}
where $(x_1,x_2)$ is the location of the target in $2$-dimensional Cartesian coordinates, $(\dot{x}_1,\dot{x}_2)$ is the corresponding speed, and $\omega$ is the turn rate. The dynamic model is
\begin{equation}
 \vx_{k+1} = \begin{pmatrix}1 & 0 & \frac{\sin(\omega_k\,\Delta t)}{\omega_k} & \frac{\cos(\omega_k\,\Delta t) - 1}{\omega_k} & 0  \\
                            0 & 1 & -\frac{\cos(\omega_k\,\Delta t) - 1}{\omega_k} & \frac{\sin(\omega_k\,\Delta t)}{\omega_k} & 0 \\ 
                            0 & 0 & \cos(\omega_k\,\Delta t) & -\sin(\omega_k\,\Delta t)& 0\\
                            0 & 0 & \sin(\omega_k\,\Delta t) & \cos(\omega_k\,\Delta t) & 0\\
                            0 & 0 & 0                      & 0                      & 1\\
             \end{pmatrix} \vx_k + \vq_k.
\end{equation}
The process noise is $\vq_k \sim \N(\vect{0},\MQ)$, where
\begin{equation}
 \MQ = \begin{pmatrix} q_\rc\,\Delta t^3 / 3 & 0               & q_\rc\,\Delta t^2 / 2 & 0                & 0  \\
                       0                 & q_\rc\,\Delta t^3/3 & 0                 & q_\rc\, \Delta t^2/2 & 0  \\
                       q_\rc\,\Delta t^2 / 2 & 0               & q_\rc\,\Delta t       & 0                & 0  \\
                       0                 & q_\rc\,\Delta t^2/2 & 0                 & q_\rc\, \Delta t     & 0  \\
                       0                 & 0                 & 0                 & 0                  & q_\omega\, \Delta t
        \end{pmatrix}.
\end{equation}
The measurements are angles from the sensors with additive Gaussian noise:
\begin{equation}
  \vy_k = \vh(\vx_k) + \vr_k,
\end{equation}
where the measurement noise is $\vr_k \sim \N(\vect{0},\MR)$. The covariance matrix $\MR$ is naturally assumed diagonal, as the measurement errors of separate sensors should be independent. For each sensor $i$ at location $\vs_{i}$ the measurement is given by
\begin{equation}
  h_i(\vx_k) = \mathrm{atan}2\!\left(x_{2,k}-s_{2,i}, x_{1,k}-s_{1,i}\right),
\end{equation}
where $\mathrm{atan}2$ is the four-quadrant inverse tangent. We focus on estimating the measurement noise variances while keeping other parameters fixed. That is, the sensor locations and dynamic model covariance are assumed to be known and the initial state distribution fixed. 

The parameters of the process noise covariance were set to $q_\rc = 0.1,~q_\omega=0.1$ and the time step to $\Delta t=0.01$. The ground-truth measurement noise covariance was $\MR = \diag(0.05^2,~ 0.1^2)$. The two sensors were located at $\vs_1 = (-1,0.5)$ and $\vs_2 = (1,1)$. The parameters of the initial distribution were $\vm_0 = (2,0,0,0,0)^\T$ and $\MP_0 = \diag(0.5^2~ 0.5^2,~ 0.5^2,~ 0.5^2,~1^2)$.  We simulated $100$ different trajectories with $T=50$ timesteps from this model. 

To compare performance of the different sigma-point schemes, we performed direct maximum likelihood estimation of the sensor noise standard deviation of the first sensor, keeping the noise of the second sensor as well as other parameters fixed at their ground truth values. The sigma-point schemes used were UKF~3, UKF~5, and UKF~7, as well as GHKF~3, GHKF~5, and GHKF~7. The $9^{\text{th}}$ order schemes were omitted since the number of sigma-points is already quite high as the state is $5$-dimensional. In addition to the sigma-point methods, we also compared against maximum likelihood estimation based on the extended Kalman filter (EKF, see, \eg\, \cite{Jazwinski:1970}).

The optimization was performed with gradient-based optimization using the Matlab optimization toolbox\footnote{MATLAB version R2014b, the \texttt{fminunc} function, quasi-Newton algorithm, initialized with $\sqrt{\MR_{1,1}}=0.1$.}. Furthermore, we investigated how the estimation performance varies as a function of uncertainty of the target's initial location. This was done by using an additional parameter for the per-coordinate standard deviation ($\sigma \in (0,0.5]$). The first two diagonal components of $\MP_0$ were set to $\sigma^2$. Furthermore, the first two components of $\vm_0$ were interpolated between the original $\vm_0$ and the simulated $\vx_0$ to keep the uncertainty of the initial location consistent with the prior.

\begin{figure}[!t]
  \pgfplotsset{
    compat=newest,    
    clip=false,
    trim axis right, trim axis left,
    ylabel style={align=center, text width=6.5cm},%
    xlabel style={xshift=3em},
    tick align=inside,
    minor tick num=1,
    legend style={font=\scriptsize, xshift=-0.25\columnwidth},
  }
  %
  %\tikzsetnextfilename{fig06}
  \setlength{\figurewidth}{0.8\columnwidth}
  \setlength{\figureheight}{0.6\columnwidth}
  \centering \sffamily \scriptsize \hspace*{4em}%
  % This file was created by matlab2tikz.
% Minimal pgfplots version: 1.3
%
%The latest updates can be retrieved from
%  http://www.mathworks.com/matlabcentral/fileexchange/22022-matlab2tikz
%where you can also make suggestions and rate matlab2tikz.
%
\definecolor{mycolor1}{rgb}{0.00000,0.44700,0.74100}%
\definecolor{mycolor2}{rgb}{0.85000,0.32500,0.09800}%
\definecolor{mycolor3}{rgb}{0.92900,0.69400,0.12500}%
\definecolor{mycolor4}{rgb}{0.49400,0.18400,0.55600}%
\definecolor{mycolor5}{rgb}{0.46600,0.67400,0.18800}%
\definecolor{mycolor6}{rgb}{0.30100,0.74500,0.93300}%
\begin{tikzpicture}

\begin{axis}[%
width=0.95092\figurewidth,
height=\figureheight,
at={(0\figurewidth,0\figureheight)},
scale only axis,
xmin=-0.1,
xmax=0.55,
xtick={0.1,0.2,0.3,0.4,0.5},
xticklabels={{$0.1$},{$0.2$},{$0.3$},{$0.4$},{$0.5$}},
xlabel={Initial location sd per coordinate},
ymode=log,
ytick={0.00001,0.0000001,0.000000001},
yticklabels={{$10^{-5}$},{$10^{-7}$},{$10^{-9}$}},
ymin=-8.85926193066376e-05,
ymax=0.000332141507179329,
yminorticks=true,
ylabel={Median absolute error of MLE $\sqrt{R_{1,1}}$ compared to GHKF 7}
]
\addplot [color=mycolor1,solid,mark=x,mark options={solid},forget plot]
  table[row sep=crcr]{%
0.5	0.000188633577782239\\
0.45	0.000125888174170374\\
0.4	9.1302648847439e-05\\
0.35	6.07783472445743e-05\\
0.3	4.13683926547177e-05\\
0.25	2.9955573798153e-05\\
0.2	2.1213256619395e-05\\
0.15	1.663998207109e-05\\
0.1	1.2133368212465e-05\\
0.05	1.25350206119992e-05\\
};
\node[above, right, align=left, inner sep=0mm, text=mycolor1]
at (axis cs:-0.05,1.25350206119992e-05,0) { EKF};
\addplot [color=mycolor2,solid,mark=x,mark options={solid},forget plot]
  table[row sep=crcr]{%
0.5	0.000113735546436709\\
0.45	7.90296625900883e-05\\
0.4	4.96228405570849e-05\\
0.35	3.94368862969806e-05\\
0.3	2.78254619141098e-05\\
0.25	1.68675276279485e-05\\
0.2	1.03825668856736e-05\\
0.15	5.67589008582808e-06\\
0.1	3.85249935825824e-06\\
0.05	2.18624576228676e-06\\
};
\node[above, right, align=left, inner sep=0mm, text=mycolor2]
at (axis cs:-0.05,2.18624576228676e-06,0) { UKF 3};
\addplot [color=mycolor3,solid,mark=x,mark options={solid},forget plot]
  table[row sep=crcr]{%
0.5	2.54445375694155e-05\\
0.45	2.06749796885126e-05\\
0.4	7.89739051996335e-06\\
0.35	2.43714998150388e-06\\
0.3	9.37379075034717e-07\\
0.25	3.35666958724973e-07\\
0.2	1.14630377679587e-07\\
0.15	3.33051125740724e-08\\
0.1	1.88766506765969e-08\\
0.05	1.55464504915381e-08\\
};
\node[above, right, align=left, inner sep=0mm, text=mycolor3]
at (axis cs:-0.05,1.55464504915381e-08,0) {};
\addplot [color=mycolor4,solid,mark=x,mark options={solid},forget plot]
  table[row sep=crcr]{%
0.5	4.08464468012772e-05\\
0.45	3.11945256895388e-05\\
0.4	9.53068009367994e-06\\
0.35	4.11763934867956e-06\\
0.3	1.1300603830651e-06\\
0.25	2.44036398720782e-07\\
0.2	5.21046002732728e-08\\
0.15	1.35794955662127e-08\\
0.1	7.07253691029086e-09\\
0.05	9.21695627459362e-09\\
};
\node[above, right, align=left, inner sep=0mm, text=mycolor4]
at (axis cs:-0.05,9.21695627459362e-09,0) { UKF 7};
\addplot [color=mycolor5,solid,mark=x,mark options={solid},forget plot]
  table[row sep=crcr]{%
0.5	2.54428386197783e-05\\
0.45	2.06758753962324e-05\\
0.4	7.89809542192307e-06\\
0.35	2.43935025170255e-06\\
0.3	9.37413776790075e-07\\
0.25	3.3395045796053e-07\\
0.2	1.13327119851542e-07\\
0.15	3.28676755985413e-08\\
0.1	1.92684782435171e-08\\
0.05	1.59165949760454e-08\\
};
\node[above, right, align=left, inner sep=0mm, text=mycolor5]
at (axis cs:-0.05,1.59165949760454e-08,0) { GHKF 3};
\addplot [color=mycolor6,solid,mark=x,mark options={solid},forget plot]
  table[row sep=crcr]{%
0.5	1.36462093519897e-05\\
0.45	1.13059647218264e-05\\
0.4	2.94024682082408e-06\\
0.35	6.86280987267918e-07\\
0.3	3.77605613409437e-08\\
0.25	6.76464276955224e-09\\
0.2	2.80312787218073e-09\\
0.15	1.94574862819041e-09\\
0.1	1.49761143070082e-09\\
0.05	1.69966917970554e-09\\
};
\node[above, right, align=left, inner sep=0mm, text=mycolor6]
at (axis cs:-0.05,1.69966917970554e-09,0) { GHKF 5};
\end{axis}
\end{tikzpicture}%
  \caption{Median absolute error of the parameter estimates compared to GHKF~7 (median taken over the $100$ simulated trajectories) as a function of the initial location prior standard deviation. UKF~5 is essentially indistinguishable from GHKF~3.}
  \label{fig:mae-of-mle}
\end{figure}
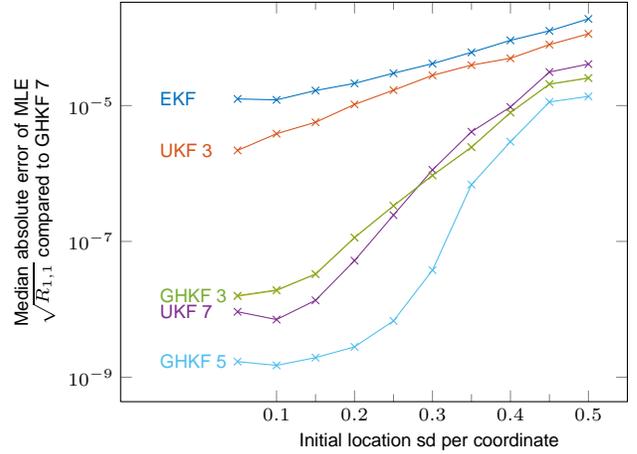

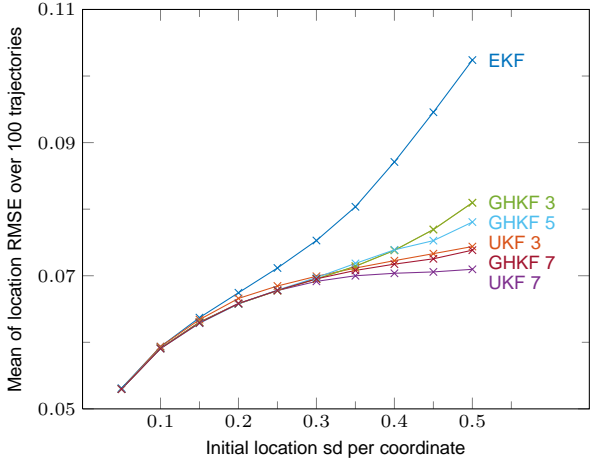
\begin{figure}[!t]
  \pgfplotsset{
    compat=newest,    
    clip=false,
    trim axis right, trim axis left,
    tick align=inside,
    ylabel style={align=center},%
    minor tick num=1,
    legend style={font=\scriptsize, xshift=-0.25\columnwidth},
  }
  %
  %\tikzsetnextfilename{fig07}
  \setlength{\figurewidth}{0.8\columnwidth}
  \setlength{\figureheight}{0.6\columnwidth}
  \centering \sffamily \scriptsize \hspace*{4em}%
  % This file was created by matlab2tikz.
% Minimal pgfplots version: 1.3
%
%The latest updates can be retrieved from
%  http://www.mathworks.com/matlabcentral/fileexchange/22022-matlab2tikz
%where you can also make suggestions and rate matlab2tikz.
%
\definecolor{mycolor1}{rgb}{0.00000,0.44700,0.74100}%
\definecolor{mycolor2}{rgb}{0.85000,0.32500,0.09800}%
\definecolor{mycolor3}{rgb}{0.92900,0.69400,0.12500}%
\definecolor{mycolor4}{rgb}{0.49400,0.18400,0.55600}%
\definecolor{mycolor5}{rgb}{0.46600,0.67400,0.18800}%
\definecolor{mycolor6}{rgb}{0.30100,0.74500,0.93300}%
\definecolor{mycolor7}{rgb}{0.63500,0.07800,0.18400}%
\begin{tikzpicture}

\begin{axis}[%
width=0.95092\figurewidth,
height=\figureheight,
at={(0\figurewidth,0\figureheight)},
scale only axis,
xmin=0,
xmax=0.65,
xtick={0.1,0.2,0.3,0.4,0.5},
xticklabels={{$0.1$},{$0.2$},{$0.3$},{$0.4$},{$0.5$}},
xlabel={Initial location sd per coordinate},
ymin=0.05,
ymax=0.11,
ytick={0.05,0.07,0.09,0.11},
yticklabels={{$0.05$},{$0.07$},{$0.09$},{$0.11$}},
ylabel={Mean of location RMSE over 100 trajectories}
]
\addplot [color=mycolor1,solid,mark=x,mark options={solid},forget plot]
  table[row sep=crcr]{%
0.5	0.102399792877139\\
0.45	0.0945642011370427\\
0.4	0.0870927926442286\\
0.35	0.0803573692674011\\
0.3	0.07528824017804\\
0.25	0.0711632600214772\\
0.2	0.067441364405894\\
0.15	0.0637051099502668\\
0.1	0.0593435693754224\\
0.05	0.0531086424886713\\
};
\node[right, align=left, inner sep=0mm, text=mycolor1]
at (axis cs:0.52,0.102399792877139,0) {EKF};
\addplot [color=mycolor2,solid,mark=x,mark options={solid},forget plot]
  table[row sep=crcr]{%
0.5	0.0743826980505496\\
0.45	0.0733154758538603\\
0.4	0.0722646844957768\\
0.35	0.0711538785638752\\
0.3	0.0699212204834705\\
0.25	0.0684587340163572\\
0.2	0.0665905517653909\\
0.15	0.0634102085985077\\
0.1	0.0593606389807995\\
0.05	0.052964685564108\\
};
\node[right, align=left, inner sep=0mm, text=mycolor2]
at (axis cs:0.52,0.075,0) {UKF 3};
\addplot [color=mycolor3,solid,mark=x,mark options={solid},forget plot]
  table[row sep=crcr]{%
0.5	0.0809669274291412\\
0.45	0.0769410136243746\\
0.4	0.0738163335930777\\
0.35	0.0714227490316147\\
0.3	0.0695166489989464\\
0.25	0.0677741654461834\\
0.2	0.0657873080458472\\
0.15	0.0630082335205575\\
0.1	0.0591128184384469\\
0.05	0.0529813029332436\\
};
\node[right, align=left, inner sep=0mm, text=mycolor3]
at (axis cs:0.52,0.0809669274291412,0) {};
\addplot [color=mycolor4,solid,mark=x,mark options={solid},forget plot]
  table[row sep=crcr]{%
0.5	0.0709647350181917\\
0.45	0.0705738517129423\\
0.4	0.0703687926585643\\
0.35	0.0699958495423029\\
0.3	0.0691547202522076\\
0.25	0.067776317534878\\
0.2	0.0658410295618871\\
0.15	0.063011979071428\\
0.1	0.0591054769962011\\
0.05	0.0529767425436818\\
};
\node[right, align=left, inner sep=0mm, text=mycolor4]
at (axis cs:0.52,0.069,0) {UKF 7};
\addplot [color=mycolor5,solid,mark=x,mark options={solid},forget plot]
  table[row sep=crcr]{%
0.5	0.080966927555545\\
0.45	0.0769410123543945\\
0.4	0.0738163318337863\\
0.35	0.0714227489211018\\
0.3	0.0695166492024582\\
0.25	0.067774165207088\\
0.2	0.0657873078599132\\
0.15	0.0630082333373733\\
0.1	0.0591128184009022\\
0.05	0.0529813028723321\\
};
\node[right, align=left, inner sep=0mm, text=mycolor5]
at (axis cs:0.52,0.081,0) {GHKF 3};
\addplot [color=mycolor6,solid,mark=x,mark options={solid},forget plot]
  table[row sep=crcr]{%
0.5	0.0780632237838428\\
0.45	0.0752704054008361\\
0.4	0.0738903250297468\\
0.35	0.07184985727229\\
0.3	0.0697217049653432\\
0.25	0.06784755237178\\
0.2	0.0657719114344536\\
0.15	0.0628571822846391\\
0.1	0.0590528331060457\\
0.05	0.0529748367784871\\
};
\node[right, align=left, inner sep=0mm, text=mycolor6]
at (axis cs:0.52,0.078,0) {GHKF 5};
\addplot [color=mycolor7,solid,mark=x,mark options={solid},forget plot]
  table[row sep=crcr]{%
0.5	0.073838644410322\\
0.45	0.0725441668202445\\
0.4	0.071742054130262\\
0.35	0.0707971033969769\\
0.3	0.0695303329347214\\
0.25	0.0677992732514066\\
0.2	0.0658206990348531\\
0.15	0.0629082771422662\\
0.1	0.0590512120527139\\
0.05	0.0529718000117662\\
};
\node[right, align=left, inner sep=0mm, text=mycolor7]
at (axis cs:0.52,0.072,0) {GHKF 7};
\end{axis}
\end{tikzpicture}%
  \caption{Mean RMSE of smoothed location (mean taken over the $100$ trajectories) using the MLE estimated noise variance of the first sensor. UKF~5 is essentially indistinguishable from GHKF~3.}
  \label{fig:rmse}
\end{figure}

Since GHKF~7 is the highest-order sigma-point scheme amongst those used in this experiment, we assume it is the most accurate and compare against it. Figure~\ref{fig:mae-of-mle} shows comparison of median (over the $100$ trajectories) absolute deviation of the MLE estimates obtained by the various filtering schemes compared to the ones obtained by GHKF~7. GHKF~5 is closest, while EKF and UKF~3 are farthest from the baseline. UKF~7, GHKF~3 and UKF~5 have similar performance. UKF~5 and GHKF~3 are essentially identical. This is explained by the observation that in 5 dimensions, all UKF~5 sigma-points are present in the GHKF~3 sigma-point set and the sum of the GHKF~3 weights of these points is $0.79$. The contribution of the remaining sigma-points that have total $0.21$ weight apparently has a negligible contribution at least with this model. 
In addition, we also look at track estimation errors using the final parameter estimates by each sigma-point scheme. Figure~\ref{fig:rmse} shows the mean RMSE over the $100$ trajectories, that is, for each simulated trajectory, we computed the smoother RMSE and then took the average.

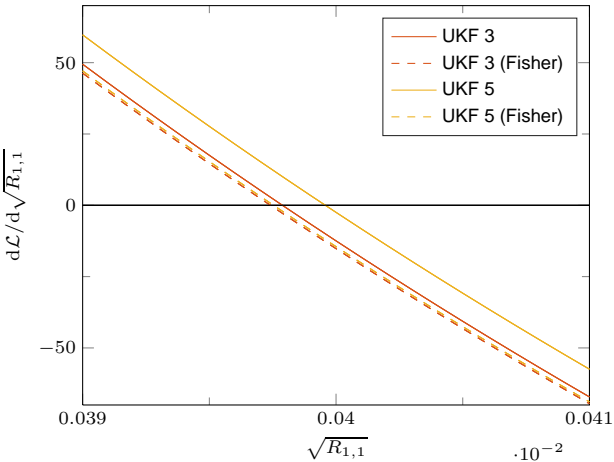
\begin{figure}[!t]
  \pgfplotsset{
    compat=newest,    
    clip=false,
    trim axis right, trim axis left,
    tick align=inside,
    ylabel style={align=center},%
    minor tick num=1,
    legend style={font=\scriptsize, xshift=-0.25\columnwidth, at={(0.92,0.97)}, anchor=north west},
  }
  %
  %\tikzsetnextfilename{fig08}
  \setlength{\figurewidth}{0.8\columnwidth}
  \setlength{\figureheight}{0.6\columnwidth}
  \centering \sffamily \scriptsize \hspace*{4em}%
  % This file was created by matlab2tikz.
% Minimal pgfplots version: 1.3
%
%The latest updates can be retrieved from
%  http://www.mathworks.com/matlabcentral/fileexchange/22022-matlab2tikz
%where you can also make suggestions and rate matlab2tikz.
%
\definecolor{mycolor1}{rgb}{0.85000,0.32500,0.09800}%
\definecolor{mycolor2}{rgb}{0.92900,0.69400,0.12500}%
\begin{tikzpicture}

\begin{axis}[%
width=0.95092\figurewidth,
height=\figureheight,
at={(0\figurewidth,0\figureheight)},
scale only axis,
xmin=0.039,
xmax=0.041,
xtick={0.039,0.04,0.041},
xticklabels={{$$$0.039$$$},{$$$0.04$$$},{$$$0.041$$$}},
xlabel={$\sqrt{R_{1,1}}$},
ymin=-70,
ymax=70,
ylabel={$\mathrm{d}\mathcal{L} / \mathrm{d}\sqrt{R_{1,1}}$},
legend style={legend cell align=left,align=left,draw=white!15!black}
]
\addplot [color=mycolor1,solid]
  table[row sep=crcr]{%
0.039	49.4582081234115\\
0.0391	42.9123262846239\\
0.0392	36.4478218489764\\
0.0393	30.0635954478849\\
0.0394	23.758564709371\\
0.0395	17.5316641906704\\
0.0396	11.3818449073402\\
0.0397	5.3080740851217\\
0.0398	-0.690665123879484\\
0.0399	-6.6153739106857\\
0.04	-12.4670381308956\\
0.0401	-18.2466284949435\\
0.0402	-23.9551009603817\\
0.0403	-29.5933967730228\\
0.0404	-35.1624431310591\\
0.0405	-40.6631526237906\\
0.0406	-46.0964248462508\\
0.0407	-51.4631449724358\\
0.0408	-56.7641855498902\\
0.0409	-62.0004050725523\\
0.041	-67.1726502749513\\
};
\addlegendentry{UKF 3};

\addplot [color=mycolor1,dashed]
  table[row sep=crcr]{%
0.039	46.3187692475838\\
0.0391	39.8150429637578\\
0.0392	33.3922702999562\\
0.0393	27.0493569336043\\
0.0394	20.785225587995\\
0.0395	14.5988157255592\\
0.0396	8.4890832512483\\
0.0397	2.45500021959515\\
0.0398	-3.50444545156392\\
0.0399	-9.39025026168179\\
0.04	-15.2033954025831\\
0.0401	-20.9448470293107\\
0.0402	-26.6155565221127\\
0.0403	-32.2164607488528\\
0.0404	-37.7484823137438\\
0.0405	-43.2125298119158\\
0.0406	-48.6094980691153\\
0.0407	-53.9402683814751\\
0.0408	-59.2057087521121\\
0.0409	-64.4066741183985\\
0.041	-69.5440065789389\\
};
\addlegendentry{UKF 3 (Fisher)};

\addplot [color=mycolor2,solid]
  table[row sep=crcr]{%
0.039	59.7480180865881\\
0.0391	53.1715977117186\\
0.0392	46.6767667898205\\
0.0393	40.2624241777907\\
0.0394	33.9274857233238\\
0.0395	27.6708841123471\\
0.0396	21.4915685409796\\
0.0397	15.3885044277986\\
0.0398	9.36067313905531\\
0.0399	3.40707146371077\\
0.04	-2.47328767709715\\
0.0401	-8.28137730144485\\
0.0402	-14.0181549113675\\
0.0403	-19.6845636048065\\
0.0404	-25.2815320572348\\
0.0405	-30.809975149518\\
0.0406	-36.2707935492484\\
0.0407	-41.664874550918\\
0.0408	-46.9930920738141\\
0.0409	-52.256306513538\\
0.041	-57.4553656600802\\
};
\addlegendentry{UKF 5};

\addplot [color=mycolor2,dashed]
  table[row sep=crcr]{%
0.039	47.1111894806502\\
0.0391	40.5993078570452\\
0.0392	34.1684660730739\\
0.0393	27.8175686899456\\
0.0394	21.545537328687\\
0.0395	15.351310366648\\
0.0396	9.23384264083188\\
0.0397	3.19210515145437\\
0.0398	-2.77491522170499\\
0.0399	-8.66821600245567\\
0.04	-14.4887793922683\\
0.0401	-20.2375725389161\\
0.0402	-25.9155478042778\\
0.0403	-31.5236430206483\\
0.0404	-37.0627817458162\\
0.0405	-42.5338735126568\\
0.0406	-47.9378140716049\\
0.0407	-53.2754856317649\\
0.0408	-58.5477570936655\\
0.0409	-63.7554842818647\\
0.041	-68.8995101655039\\
};
\addlegendentry{UKF 5 (Fisher)};

\addplot [color=black,solid,forget plot]
  table[row sep=crcr]{%
0.039	0\\
0.041	0\\
};
\addplot [color=mycolor1,solid,forget plot]
  table[row sep=crcr]{%
0.039	49.4582081234115\\
0.0391	42.9123262846239\\
0.0392	36.4478218489764\\
0.0393	30.0635954478849\\
0.0394	23.758564709371\\
0.0395	17.5316641906704\\
0.0396	11.3818449073402\\
0.0397	5.3080740851217\\
0.0398	-0.690665123879484\\
0.0399	-6.6153739106857\\
0.04	-12.4670381308956\\
0.0401	-18.2466284949435\\
0.0402	-23.9551009603817\\
0.0403	-29.5933967730228\\
0.0404	-35.1624431310591\\
0.0405	-40.6631526237906\\
0.0406	-46.0964248462508\\
0.0407	-51.4631449724358\\
0.0408	-56.7641855498902\\
0.0409	-62.0004050725523\\
0.041	-67.1726502749513\\
};
\addplot [color=mycolor1,dashed,forget plot]
  table[row sep=crcr]{%
0.039	46.3187692475838\\
0.0391	39.8150429637578\\
0.0392	33.3922702999562\\
0.0393	27.0493569336043\\
0.0394	20.785225587995\\
0.0395	14.5988157255592\\
0.0396	8.4890832512483\\
0.0397	2.45500021959515\\
0.0398	-3.50444545156392\\
0.0399	-9.39025026168179\\
0.04	-15.2033954025831\\
0.0401	-20.9448470293107\\
0.0402	-26.6155565221127\\
0.0403	-32.2164607488528\\
0.0404	-37.7484823137438\\
0.0405	-43.2125298119158\\
0.0406	-48.6094980691153\\
0.0407	-53.9402683814751\\
0.0408	-59.2057087521121\\
0.0409	-64.4066741183985\\
0.041	-69.5440065789389\\
};
\addplot [color=mycolor2,solid,forget plot]
  table[row sep=crcr]{%
0.039	59.7480180865881\\
0.0391	53.1715977117186\\
0.0392	46.6767667898205\\
0.0393	40.2624241777907\\
0.0394	33.9274857233238\\
0.0395	27.6708841123471\\
0.0396	21.4915685409796\\
0.0397	15.3885044277986\\
0.0398	9.36067313905531\\
0.0399	3.40707146371077\\
0.04	-2.47328767709715\\
0.0401	-8.28137730144485\\
0.0402	-14.0181549113675\\
0.0403	-19.6845636048065\\
0.0404	-25.2815320572348\\
0.0405	-30.809975149518\\
0.0406	-36.2707935492484\\
0.0407	-41.664874550918\\
0.0408	-46.9930920738141\\
0.0409	-52.256306513538\\
0.041	-57.4553656600802\\
};
\addplot [color=mycolor2,dashed,forget plot]
  table[row sep=crcr]{%
0.039	47.1111894806502\\
0.0391	40.5993078570452\\
0.0392	34.1684660730739\\
0.0393	27.8175686899456\\
0.0394	21.545537328687\\
0.0395	15.351310366648\\
0.0396	9.23384264083188\\
0.0397	3.19210515145437\\
0.0398	-2.77491522170499\\
0.0399	-8.66821600245567\\
0.04	-14.4887793922683\\
0.0401	-20.2375725389161\\
0.0402	-25.9155478042778\\
0.0403	-31.5236430206483\\
0.0404	-37.0627817458162\\
0.0405	-42.5338735126568\\
0.0406	-47.9378140716049\\
0.0407	-53.2754856317649\\
0.0408	-58.5477570936655\\
0.0409	-63.7554842818647\\
0.041	-68.8995101655039\\
};
\addplot [color=black,solid,forget plot]
  table[row sep=crcr]{%
0.039	0\\
0.041	0\\
};
\addplot [color=mycolor1,solid,forget plot]
  table[row sep=crcr]{%
0.039	49.4582081234115\\
0.0391	42.9123262846239\\
0.0392	36.4478218489764\\
0.0393	30.0635954478849\\
0.0394	23.758564709371\\
0.0395	17.5316641906704\\
0.0396	11.3818449073402\\
0.0397	5.3080740851217\\
0.0398	-0.690665123879484\\
0.0399	-6.6153739106857\\
0.04	-12.4670381308956\\
0.0401	-18.2466284949435\\
0.0402	-23.9551009603817\\
0.0403	-29.5933967730228\\
0.0404	-35.1624431310591\\
0.0405	-40.6631526237906\\
0.0406	-46.0964248462508\\
0.0407	-51.4631449724358\\
0.0408	-56.7641855498902\\
0.0409	-62.0004050725523\\
0.041	-67.1726502749513\\
};
\addplot [color=mycolor1,dashed,forget plot]
  table[row sep=crcr]{%
0.039	46.3187692475838\\
0.0391	39.8150429637578\\
0.0392	33.3922702999562\\
0.0393	27.0493569336043\\
0.0394	20.785225587995\\
0.0395	14.5988157255592\\
0.0396	8.4890832512483\\
0.0397	2.45500021959515\\
0.0398	-3.50444545156392\\
0.0399	-9.39025026168179\\
0.04	-15.2033954025831\\
0.0401	-20.9448470293107\\
0.0402	-26.6155565221127\\
0.0403	-32.2164607488528\\
0.0404	-37.7484823137438\\
0.0405	-43.2125298119158\\
0.0406	-48.6094980691153\\
0.0407	-53.9402683814751\\
0.0408	-59.2057087521121\\
0.0409	-64.4066741183985\\
0.041	-69.5440065789389\\
};
\addplot [color=mycolor2,solid,forget plot]
  table[row sep=crcr]{%
0.039	59.7480180865881\\
0.0391	53.1715977117186\\
0.0392	46.6767667898205\\
0.0393	40.2624241777907\\
0.0394	33.9274857233238\\
0.0395	27.6708841123471\\
0.0396	21.4915685409796\\
0.0397	15.3885044277986\\
0.0398	9.36067313905531\\
0.0399	3.40707146371077\\
0.04	-2.47328767709715\\
0.0401	-8.28137730144485\\
0.0402	-14.0181549113675\\
0.0403	-19.6845636048065\\
0.0404	-25.2815320572348\\
0.0405	-30.809975149518\\
0.0406	-36.2707935492484\\
0.0407	-41.664874550918\\
0.0408	-46.9930920738141\\
0.0409	-52.256306513538\\
0.041	-57.4553656600802\\
};
\addplot [color=mycolor2,dashed,forget plot]
  table[row sep=crcr]{%
0.039	47.1111894806502\\
0.0391	40.5993078570452\\
0.0392	34.1684660730739\\
0.0393	27.8175686899456\\
0.0394	21.545537328687\\
0.0395	15.351310366648\\
0.0396	9.23384264083188\\
0.0397	3.19210515145437\\
0.0398	-2.77491522170499\\
0.0399	-8.66821600245567\\
0.04	-14.4887793922683\\
0.0401	-20.2375725389161\\
0.0402	-25.9155478042778\\
0.0403	-31.5236430206483\\
0.0404	-37.0627817458162\\
0.0405	-42.5338735126568\\
0.0406	-47.9378140716049\\
0.0407	-53.2754856317649\\
0.0408	-58.5477570936655\\
0.0409	-63.7554842818647\\
0.041	-68.8995101655039\\
};
\addplot [color=black,solid,forget plot]
  table[row sep=crcr]{%
0.039	0\\
0.041	0\\
};
\end{axis}
\end{tikzpicture}%
  \caption{Derivative of the log-likelihood with respect to the standard deviation of the error of the first sensor. Evaluated using UKF 3 and UKF 5 both with the sensitivity equation approach (Section \ref{sec:direct}) and the Fisher identity based approach (Section \ref{sec:fisher}).}
  \label{fig:ct_Fishercomparison}
\end{figure}

To compare the two different gradient evaluation approaches, sensitivity equations (Section \ref{sec:direct}) and the Fisher identity approach (Section \ref{sec:fisher}), we evaluated the derivative of the log-likelihood with respect to the standard deviation of the error of the first sensor, using UKF 3 and UKF 5 and both gradient evaluation approaches. The results are shown in Figure~\ref{fig:ct_Fishercomparison}. With UKF 3, there is clear difference between the estimated gradients while with UKF 5 the approaches essentially agree.

\begin{figure}[!t]
  \pgfplotsset{
    compat=newest,    
    clip=false,
    trim axis right, trim axis left,
    tick align=inside,
    ylabel style={align=center},%
    minor tick num=1,
    legend style={font=\scriptsize, xshift=-0.25\columnwidth, at={(0.37,0.05)},anchor=south west},
  }
  %
  %\tikzsetnextfilename{fig09}
  \setlength{\figurewidth}{0.8\columnwidth}
  \setlength{\figureheight}{0.6\columnwidth}
  \centering \sffamily \scriptsize \hspace*{4em}%
  \input{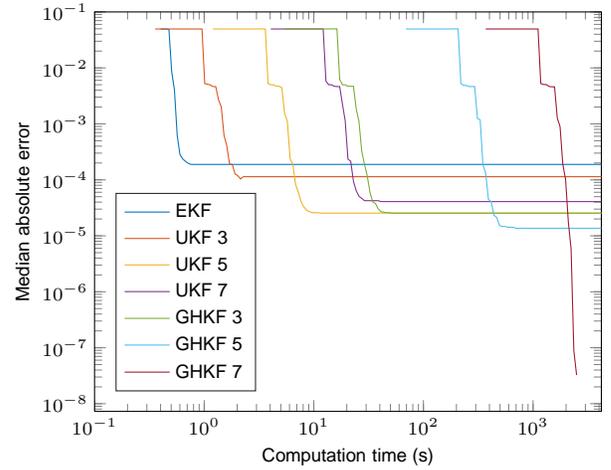}
  \caption{Median absolute error of the parameter as a function of computation time during the optimization. Median taken over 100 datasets.}
  \label{fig:comptime}
\end{figure}

\begin{figure}[!t]
  \pgfplotsset{
    compat=newest,    
    clip=false,
    trim axis right, trim axis left,
    tick align=inside,
    ylabel style={align=center},%
    minor tick num=1,
    legend style={font=\scriptsize, xshift=-0.25\columnwidth, at={(1.05,0.965)}, anchor=north west},
  }
  %
  %\tikzsetnextfilename{fig10}
  \setlength{\figurewidth}{0.8\columnwidth}
  \setlength{\figureheight}{0.6\columnwidth}
  \centering \sffamily \scriptsize \hspace*{4em}%
  \input{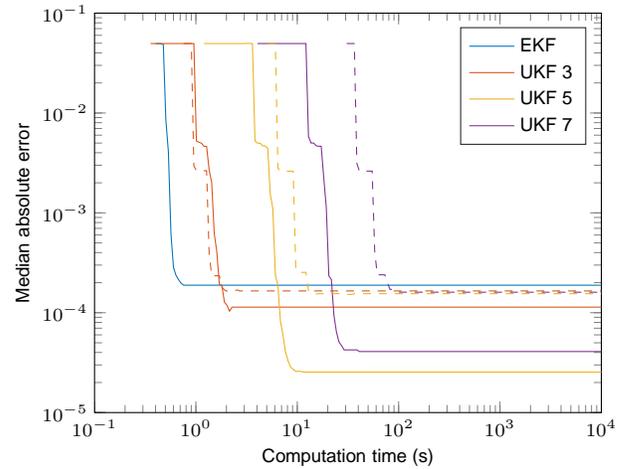}
  \caption{Median absolute error of the parameter as a function of computation time during the optimization. Median taken over 100 datasets. The solid lines are gradient-based direct optimization, while the dashed lines show EM (run for 32 iterations) with the corresponding sigma-point schemes.}
  \label{fig:comptime_EM}
\end{figure}

To measure the performance as a function of computational cost, we recorded the parameter values as well as the times used at each iteration of the optimization routines. In this experiment, the initial location standard deviation per coordinate was set to $0.5$. Median absolute error (compared to final GHKF 7 estimate, as a function of time) is shown in Figures~\ref{fig:comptime} and \ref{fig:comptime_EM}. As one would expect, the higher-order schemes are more computationally demanding and GHKF is more computationally demanding than UKF, but eventually the higher-order GHKF schemes find better parameter estimates.

\begin{figure}[!t]
  \pgfplotsset{
    compat=newest,    
    clip=false,
    tick align=inside,
    y tick label style={rotate=90},
    y tick scale label style={rotate=-90},
    minor tick num=1,
    legend style={font=\scriptsize, xshift=-0.25\columnwidth},
  }
  \setlength{\figurewidth}{0.37\columnwidth}
  \setlength{\figureheight}{0.4\columnwidth}
  \centering \sffamily \scriptsize
  %\tikzsetnextfilename{fig11a}
  \subfloat{% This file was created by matlab2tikz.
% Minimal pgfplots version: 1.3
%
%The latest updates can be retrieved from
%  http://www.mathworks.com/matlabcentral/fileexchange/22022-matlab2tikz
%where you can also make suggestions and rate matlab2tikz.
%
\definecolor{mycolor1}{rgb}{0.85000,0.32500,0.09800}%
\definecolor{mycolor2}{rgb}{0.92900,0.69400,0.12500}%
\definecolor{mycolor3}{rgb}{0.49400,0.18400,0.55600}%
\begin{tikzpicture}

\begin{axis}[%
width=0.95092\figurewidth,
height=\figureheight,
at={(0\figurewidth,0\figureheight)},
scale only axis,
xmin=1,
xmax=10,
xlabel={EM step},
ymin=0.038,
ymax=0.056,
ylabel={$\sqrt{R_{1,1}}$}
]
\addplot [color=mycolor1,solid,forget plot]
  table[row sep=crcr]{%
1	0.0548255883358975\\
2	0.040638297188241\\
3	0.0397928879438972\\
4	0.0397440386604441\\
5	0.0397412174535953\\
6	0.0397410545232108\\
7	0.0397410451136636\\
8	0.039741044570244\\
9	0.0397410445388602\\
10	0.0397410445370475\\
};
\addplot [color=mycolor2,solid,forget plot]
  table[row sep=crcr]{%
1	0.0539321040953216\\
2	0.0405883901710265\\
3	0.0398013020979053\\
4	0.039756094693554\\
5	0.0397534991412695\\
6	0.039753350121847\\
7	0.039753341566149\\
8	0.039753341074938\\
9	0.0397533410467359\\
10	0.0397533410451165\\
};
\addplot [color=mycolor3,solid,forget plot]
  table[row sep=crcr]{%
1	0.0538368995070977\\
2	0.0406031779158601\\
3	0.0398228916463862\\
4	0.0397782204532634\\
5	0.0397756644088012\\
6	0.0397755181579666\\
7	0.0397755097898367\\
8	0.0397755093110553\\
9	0.0397755092836606\\
10	0.0397755092820927\\
};
\end{axis}
\end{tikzpicture}%
}%
  \hfil
  %\tikzsetnextfilename{fig11b}
  \subfloat{% This file was created by matlab2tikz.
% Minimal pgfplots version: 1.3
%
%The latest updates can be retrieved from
%  http://www.mathworks.com/matlabcentral/fileexchange/22022-matlab2tikz
%where you can also make suggestions and rate matlab2tikz.
%
\definecolor{mycolor1}{rgb}{0.85000,0.32500,0.09800}%
\definecolor{mycolor2}{rgb}{0.92900,0.69400,0.12500}%
\definecolor{mycolor3}{rgb}{0.49400,0.18400,0.55600}%
\begin{tikzpicture}

\begin{axis}[%
width=0.95092\figurewidth,
height=\figureheight,
at={(0\figurewidth,0\figureheight)},
scale only axis,
xmin=10,
xmax=100,
xlabel={EM step},
ymin=0.0397,
ymax=0.04,
ylabel={$\sqrt{R_{1,1}}$}
]
\addplot [color=mycolor1,solid,forget plot]
  table[row sep=crcr]{%
10	0.0397410445370475\\
11	0.0397410445369428\\
12	0.0397410445369367\\
13	0.0397410445369365\\
14	0.0397410445369365\\
15	0.0397410445369364\\
16	0.0397410445369365\\
17	0.0397410445369365\\
18	0.0397410445369365\\
19	0.0397410445369365\\
20	0.0397410445369365\\
21	0.0397410445369365\\
22	0.0397410445369365\\
23	0.0397410445369365\\
24	0.0397410445369365\\
25	0.0397410445369365\\
26	0.0397410445369365\\
27	0.0397410445369365\\
28	0.0397410445369365\\
29	0.0397410445369365\\
30	0.0397410445369365\\
31	0.0397410445369365\\
32	0.0397410445369365\\
33	0.0397410445369365\\
34	0.0397410445369365\\
35	0.0397410445369365\\
36	0.0397410445369365\\
37	0.0397410445369365\\
38	0.0397410445369365\\
39	0.0397410445369365\\
40	0.0397410445369365\\
41	0.0397410445369365\\
42	0.0397410445369365\\
43	0.0397410445369365\\
44	0.0397410445369365\\
45	0.0397410445369365\\
46	0.0397410445369365\\
47	0.0397410445369365\\
48	0.0397410445369365\\
49	0.0397410445369365\\
50	0.0397410445369365\\
51	0.0397410445369365\\
52	0.0397410445369365\\
53	0.0397410445369365\\
54	0.0397410445369365\\
55	0.0397410445369365\\
56	0.0397410445369365\\
57	0.0397410445369365\\
58	0.0397410445369365\\
59	0.0397410445369365\\
60	0.0397410445369365\\
61	0.0397410445369365\\
62	0.0397410445369365\\
63	0.0397410445369365\\
64	0.0397410445369365\\
65	0.0397410445369365\\
66	0.0397410445369365\\
67	0.0397410445369365\\
68	0.0397410445369365\\
69	0.0397410445369365\\
70	0.0397410445369365\\
71	0.0397410445369365\\
72	0.0397410445369365\\
73	0.0397410445369365\\
74	0.0397410445369365\\
75	0.0397410445369365\\
76	0.0397410445369365\\
77	0.0397410445369365\\
78	0.0397410445369365\\
79	0.0397410445369365\\
80	0.0397410445369365\\
81	0.0397410445369365\\
82	0.0397410445369365\\
83	0.0397410445369365\\
84	0.0397410445369365\\
85	0.0397410445369365\\
86	0.0397410445369365\\
87	0.0397410445369365\\
88	0.0397410445369365\\
89	0.0397410445369365\\
90	0.0397410445369365\\
91	0.0397410445369365\\
92	0.0397410445369365\\
93	0.0397410445369365\\
94	0.0397410445369365\\
95	0.0397410445369365\\
96	0.0397410445369365\\
97	0.0397410445369365\\
98	0.0397410445369365\\
99	0.0397410445369365\\
100	0.0397410445369365\\
};
\addplot [color=mycolor2,solid,forget plot]
  table[row sep=crcr]{%
10	0.0397533410451165\\
11	0.0397533410450234\\
12	0.0397533410450186\\
13	0.039753341045019\\
14	0.039753341045019\\
15	0.0397533410450191\\
16	0.039753341045019\\
17	0.0397533410450191\\
18	0.039753341045019\\
19	0.0397533410450192\\
20	0.0397533410450191\\
21	0.039753341045019\\
22	0.0397533410450191\\
23	0.039753341045019\\
24	0.0397533410450191\\
25	0.039753341045019\\
26	0.0397533410450191\\
27	0.039753341045019\\
28	0.0397533410450192\\
29	0.0397533410450191\\
30	0.039753341045019\\
31	0.0397533410450191\\
32	0.039753341045019\\
33	0.0397533410450191\\
34	0.039753341045019\\
35	0.0397533410450191\\
36	0.039753341045019\\
37	0.0397533410450192\\
38	0.0397533410450191\\
39	0.039753341045019\\
40	0.0397533410450191\\
41	0.039753341045019\\
42	0.0397533410450191\\
43	0.039753341045019\\
44	0.0397533410450191\\
45	0.039753341045019\\
46	0.0397533410450192\\
47	0.0397533410450191\\
48	0.039753341045019\\
49	0.0397533410450191\\
50	0.039753341045019\\
51	0.0397533410450191\\
52	0.039753341045019\\
53	0.0397533410450191\\
54	0.039753341045019\\
55	0.0397533410450192\\
56	0.0397533410450191\\
57	0.039753341045019\\
58	0.0397533410450191\\
59	0.039753341045019\\
60	0.0397533410450191\\
61	0.039753341045019\\
62	0.0397533410450191\\
63	0.039753341045019\\
64	0.0397533410450192\\
65	0.0397533410450191\\
66	0.039753341045019\\
67	0.0397533410450191\\
68	0.039753341045019\\
69	0.0397533410450191\\
70	0.039753341045019\\
71	0.0397533410450191\\
72	0.039753341045019\\
73	0.0397533410450192\\
74	0.0397533410450191\\
75	0.039753341045019\\
76	0.0397533410450191\\
77	0.039753341045019\\
78	0.0397533410450191\\
79	0.039753341045019\\
80	0.0397533410450191\\
81	0.039753341045019\\
82	0.0397533410450192\\
83	0.0397533410450191\\
84	0.039753341045019\\
85	0.0397533410450191\\
86	0.039753341045019\\
87	0.0397533410450191\\
88	0.039753341045019\\
89	0.0397533410450191\\
90	0.039753341045019\\
91	0.0397533410450192\\
92	0.0397533410450191\\
93	0.039753341045019\\
94	0.0397533410450191\\
95	0.039753341045019\\
96	0.0397533410450191\\
97	0.039753341045019\\
98	0.0397533410450191\\
99	0.039753341045019\\
100	0.0397533410450192\\
};
\addplot [color=mycolor3,solid,forget plot]
  table[row sep=crcr]{%
10	0.0397755092820927\\
11	0.039775509282008\\
12	0.0397755092820005\\
13	0.0397755092819995\\
14	0.0397755092819993\\
15	0.0397755092819991\\
16	0.0397755092819989\\
17	0.0397755092819995\\
18	0.0397755092819983\\
19	0.0397755092819982\\
20	0.0397755092819997\\
21	0.0397755092819992\\
22	0.0397755092819993\\
23	0.0397755092819991\\
24	0.0397755092819987\\
25	0.0397755092819998\\
26	0.0397755092819986\\
27	0.0397755092819994\\
28	0.0397755092819992\\
29	0.0397755092819992\\
30	0.0397755092819995\\
31	0.0397755092819992\\
32	0.0397755092819985\\
33	0.0397755092819999\\
34	0.0397755092820003\\
35	0.0397755092820002\\
36	0.0397755092820001\\
37	0.0397755092819995\\
38	0.0397755092819983\\
39	0.0397755092819996\\
40	0.0397755092819993\\
41	0.0397755092819993\\
42	0.0397755092819978\\
43	0.0397755092819985\\
44	0.0397755092819986\\
45	0.0397755092819984\\
46	0.0397755092819984\\
47	0.0397755092820001\\
48	0.0397755092820008\\
49	0.0397755092819993\\
50	0.0397755092819998\\
51	0.0397755092819986\\
52	0.0397755092819978\\
53	0.0397755092819998\\
54	0.0397755092819983\\
55	0.0397755092819977\\
56	0.039775509281998\\
57	0.0397755092819981\\
58	0.0397755092819985\\
59	0.0397755092819986\\
60	0.0397755092819984\\
61	0.0397755092819984\\
62	0.0397755092820001\\
63	0.0397755092820008\\
64	0.0397755092819993\\
65	0.0397755092819998\\
66	0.0397755092819986\\
67	0.0397755092819978\\
68	0.0397755092819998\\
69	0.0397755092819983\\
70	0.0397755092819977\\
71	0.039775509281998\\
72	0.0397755092819981\\
73	0.0397755092819985\\
74	0.0397755092819986\\
75	0.0397755092819984\\
76	0.0397755092819984\\
77	0.0397755092820001\\
78	0.0397755092820008\\
79	0.0397755092819993\\
80	0.0397755092819998\\
81	0.0397755092819986\\
82	0.0397755092819978\\
83	0.0397755092819998\\
84	0.0397755092819983\\
85	0.0397755092819977\\
86	0.039775509281998\\
87	0.0397755092819981\\
88	0.0397755092819985\\
89	0.0397755092819986\\
90	0.0397755092819984\\
91	0.0397755092819984\\
92	0.0397755092820001\\
93	0.0397755092820008\\
94	0.0397755092819993\\
95	0.0397755092819998\\
96	0.0397755092819986\\
97	0.0397755092819978\\
98	0.0397755092819998\\
99	0.0397755092819983\\
100	0.0397755092819977\\
};
\addplot [color=mycolor1,dashed,forget plot]
  table[row sep=crcr]{%
10	0.0397884192696408\\
100	0.0397884192696408\\
};
\addplot [color=mycolor2,dashed,forget plot]
  table[row sep=crcr]{%
10	0.0399577863107484\\
100	0.0399577863107484\\
};
\addplot [color=mycolor3,dashed,forget plot]
  table[row sep=crcr]{%
10	0.0399346209845639\\
100	0.0399346209845639\\
};
\node[right, align=left, inner sep=0mm, text=mycolor1]
at (axis cs:70,0.0397490445369365,0) {\tiny EM UKF 3};
\node[right, align=left, inner sep=0mm, text=mycolor2]
at (axis cs:40,0.0397613410450192,0) {\tiny EM UKF 5};
\node[right, align=left, inner sep=0mm, text=mycolor3]
at (axis cs:70,0.0397835092819977,0) {\tiny EM UKF 7};
\node[right, align=left, inner sep=0mm, text=mycolor1]
at (axis cs:70,0.0397964192696408,0) {\tiny MLE UKF 3};
\node[right, align=left, inner sep=0mm, text=mycolor2]
at (axis cs:70,0.0399657863107484,0) {\tiny MLE UKF 5};
\node[right, align=left, inner sep=0mm, text=mycolor3]
at (axis cs:70,0.0399426209845639,0) {\tiny MLE UKF 7};
\end{axis}
\end{tikzpicture}%
}%

  \caption{The plot on the left shows the evolution of the parameter estimate during the first 10 EM steps using UKF~3, UKF~5, and UKF~7. All methods converge essentially at same speed to the same value. The plot on the right shows the evolution of the parameter estimates after the first 10 EM steps as well as the corresponding direct MLE estimates (the dashed lines).}
  \label{fig:ct_100thdataset_em}
\end{figure}
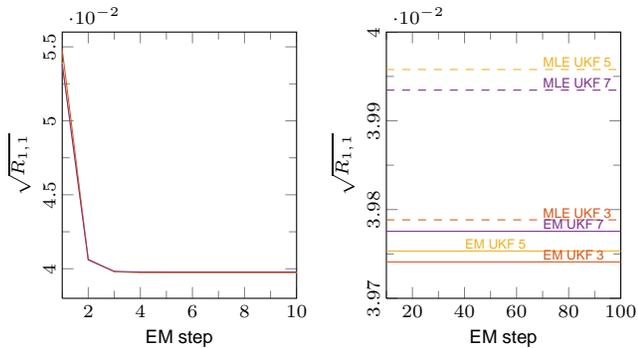

Figure~\ref{fig:ct_100thdataset_em} shows the evolution of EM parameter estimation for one simulated trajectory with $\sigma=0.5$. The EM algorithm practically converges in a couple of steps with all three sigma-point schemes, and the final parameter estimates are rather close to each other and to the direct MLE estimates. Theoretically, the EM algorithm has linear convergence \cite{Dempster:1977} although it is hard to say whether these convergence results extend to the case where the E-step is approximated using sigma-point smoothers.

\section{Conclusion and Discussion}
In this paper together with the complementing conference article \cite{Kokkala+Solin+Sarkka:2014}, we have considered various probabilistic point estimation approaches for parameter estimation in nonlinear system identification. We discussed direct likelihood maximization as well as the expectation--maximization (EM) algorithm coupled with various filtering and smoothing algorithms, namely, sigma-point filters, particle filters, and extended Kalman filters as well as the corresponding smoothers. In this paper, we focused on the differences between different sigma-point filters based on unscented transforms of third, fifth, seventh and ninth orders, and the Gauss--Hermite cubature rules.

In diminishing order of computational complexity and theoretical exactness, the filtering methods would rank as follows: particle filter, sigma-point filter, extended Kalman filter based direct likelihood approximation. In theory, particle filters converge to the exact filtering solution as the number of particles increases, while the other methods considered are based on assuming a Gaussian density and using the Kalman filter equations. However, especially in high-dimensional cases, the computational cost may prohibit the use of particle filters. In practice, the assumed density Gaussian filtering approach may have satisfactory performance if the nonlinearity is not too high. In principle, all sigma-point filters are based on assuming Gaussian density, and a higher-order cubature rule should lead to more accurate approximation of the Gaussian integrals at the cost of higher computational burden. However, typically in the literature (\eg\ \cite{Arasaratnam+Haykin:2009}) it has been claimed that when the Gaussian density approximation is already inaccurate, more accurate computation of the Gaussian integrals is not beneficial.

We also tested the methods in two simulated case studies. In the univariate nonstationary growth model, maximum likelihood estimates produced by different sigma-point schemes were similar. However, the estimates obtained by higher-order unscented schemes were closer to the Gauss--Hermite (order 16) baseline than the conventional $3^{\text{rd}}$ order unscented transform. This suggests that the higher order methods may indeed have some utility.

In the target tracking experiment, we compared the estimates of the noise standard deviation of one of the sensors as a function of prior uncertainty of the target's location using each of the sigma-point schemes. With higher prior uncertainty, there were more differences amongst the methods. This is reasonable because when there is less uncertainty in the model, all the methods obtain more accurate parameter estimates. Furthermore, the nonlinearity of the model has a stronger effect when the state variance is larger. Since no exact maximum likelihood estimate was available, we compared to the highest-order sigma-point scheme, namely, GHKF~7. Compared to that, the Gauss--Hermite schemes were closer than the unscented transform based schemes and higher order schemes were closer than lower order schemes. Thus, the results are consistent with an assumption that the higher-order sigma-point methods produce better approximations to the Gaussian filtering result.

We also measured the performance of the discussed optimization routines as a function of computational time. In direct gradient-based optimization, higher-degree algorithms are more time-consuming but eventually seem to obtain better parameter estimates. The EM algorithm with low-degree sigma-point schemes (UKF~3, UKF~5) was initially faster than the gradient-based optimization, but eventually the gradient-based optimization seems to obtain better values. The EM algorithm with UKF~7 sigma-points was more computationally demanding than the direct optimization with UKF~7. This is due to the fact that EM requires the $2n$-dimensional sigma-points for the smoothing distribution. This suggests that EM is not applicable in high-dimensional problems combined with high-degree sigma-point schemes. However, when interpreting these findings, it should be noted that we compared to the GHKF~7 estimate since the true maximum-likelihood estimate was not available. 

The sigma-point integration schemes are derived by assuming exact integration results for polynomials of certain degrees. Thus, it should be noted that it is not guaranteed that a higher-order integration rule produces a more accurate results, even though it is accurate for higher-order polynomials. Furthermore, it is not guaranteed that a better approximation to the Gaussian filtering result produces a better approximation to the exact maximum likelihood result. On the other hand, there is no reason why in general a lower-order approximation to the Gaussian filtering integrals would produce more accurate approximations to the exact filtering results.

We also compared the actual tracking performance in terms of the smoother root mean square errors of the target locations using the smoother results obtained with the maximum likelihood parameter estimate of each sigma-point filter. There was no clear differences between the different sigma-point schemes in terms of the tracking error in this experiment. The tracking error of EKF increased more rapidly as a function of the initial location uncertainty, which demonstrates the local linearization nature of EKF.

In five dimensions, the UKF~5 sigma-point scheme approximates the GHKF~3 scheme in the sense that all UKF~5 sigma-points are GHKF~3 sigma-points as well, and more than half of total weight is contributed by these points. The target-tracking experiment demonstrated that these two schemes indeed produce almost equal results. However, most GHKF~3 sigma-points are not used in the UKF~5 scheme, and thus evaluating an integral by UKF~5 requires considerably fewer function evaluations. These results suggest that there is no reason to use GHKF~3 in five dimensions as UKF~5 produces essentially same results with fewer computations. It is possible that similar computationally lighter close approximations exist for other Gauss--Hermite based sigma-point schemes as well.

In the target-tracking experiment, we also investigated how the performance of the EM algorithm varies with the sigma-point scheme used. As a function of EM iterations, the evolution of the parameter estimate was not affected by the choice of sigma-point scheme. However, the number of sigma-points in the UKF~3 rule is a small fraction of the number of sigma-points with the higher order rules. Thus, measured by model function evaluations, EM with the UKF~3 rule converged faster. This suggests that even when the interest lies in obtaining as accurate parameter estimates as possible, a reasonable computational approach would be to first use EM with a low-order sigma-point scheme, such as UKF~3, to obtain a ballpark estimate. Then, if accuracy is desired, the initial estimate could be refined using a more accurate sigma-point scheme combined with a direct optimization algorithm.

In this paper, we considered only discrete-time state-space models. Different sigma-point schemes may also be used for continuous-discrete state-space models (see, \eg, \cite{Crouse:2015} and references therein). We considered only fixed deterministic sigma-point schemes. An interesting future research topic could be to combine the recently proposed filters based on adapting or randomizing the sigma-points \cite{Straka+Dunik+Simandl+Blasch:2012,Dunik+Straka+Simandl:2013,Straka+Dunik+Simandl+Blasch:2014} with parameter estimation. 

Finally, we attempt to conclude which of the methods considered here one should use in practice. Regarding the choice of sigma-point methods the higher order unscented transform methods turned out to be quite good in the examples that we considered---but if the best possible accuracy is desired, then Gauss--Hermite methods need to be used. The EM algorithm is indeed useful in situations when the M-step optimization can be done in closed form---of which important special cases are the linear-in-parameters models considered here. However, for nonlinear in parameters models EM might not be a good choice. For nonlinear in parameters models it is thus beneficial to directly optimize the log-likelihood, and in that case we have the choice to evaluate the gradients either using the sensitivity equations or using the Fisher's identity. It turned out that the Fisher's identity is often computationally more demanding than the sensitivity equations, due to the requirement of smoothing pass, which favors the use of sensitivity equations for this purpose. Furthermore, the sensitivity equations give the exact gradients of the approximate likelihood whereas Fisher's identity only gives approximate gradients of it. However, the Fisher's identity has the advantage of easy black-box implementation which can sometimes be seen as an advantage.

\appendix[M-step in the linear-in-parameters case]
By substituting the linear-in-parameters model ($\vf(\vx):= \MA\,\tilde{\vf}(\vx),~\vh(\vx):= \MH\,\tilde{\vh}(\vx)$) into the general expression for $\mathcal{Q}(\vtheta,\vtheta^{(n)})$, we obtain
\begin{align}
  &\mathcal{Q}(\vtheta,\vtheta^{(n)}) \nonumber \\ 
  &\approx -\frac{1}{2} \log |2\pi\,\MP_0|
  - \frac{T}{2} \log |2\pi\,\MQ| 
  - \frac{T}{2} \log |2\pi\,\MR| \nonumber \\ 
  &  -\frac{1}{2} \tr \Big\{ \MP_0^{-1} \,  \Big[ \MP_{0 \mid T} + (\vm_{0 \mid T} - \vm_0) \, (\vm_{0 \mid T} - \vm_0)^\T
      \Big] \Big\} \nonumber \\ 
  &  -\frac{1}{2} \sum_{k=1}^T
     \tr \big\{ \MQ^{-1} \,  \E\big[ (\vx_{k} - \MA\,\tilde{\vf}(\vx_{k-1}) ) \nonumber\\
   &\qquad \qquad \qquad \times   (\vx_{k} - \MA\,\tilde{\vf}(\vx_{k-1}))^\T
       \mid  \vy_{1:T} \big] \big\} \nonumber \\
  &  -\frac{1}{2} \sum_{k=1}^T
     \tr \big\{ \MR^{-1} \,   \E\big[ (\vy_{k} - \MH\,\tilde{\vh}(\vx_{k})) \nonumber \\
   &\qquad \qquad \qquad \times
                   (\vy_{k} - \MH\,\tilde{\vh}(\vx_{k}))^\T
       \mid  \vy_{1:T} \big] \big\}.
\label{eq:appendixQ}       
\end{align}
Since trace is linear, the penultimate term can be written as
\begin{align}
=& - \frac{T}{2} \tr \big\{ \MQ^{-1}\, \frac{1}{T}\, \sum_{k=1}^T
  \E\big[ (\vx_{k} - \MA\,\tilde{\vf}(\vx_{k-1}) ) \nonumber \\
   &\qquad \qquad \qquad \qquad \times
                   (\vx_{k} - \MA\,\tilde{\vf}(\vx_{k-1}))^\T
       \mid  \vy_{1:T} \big] \big\} \nonumber \\ 
=& - \frac{T}{2} \tr \big\{ \MQ^{-1} \,\frac{1}{T}\,\sum_{k=1}^T \E\big[ \vx_k\,\vx_k^\T - \vx_k\,\tilde{\vf}(\vx_{k-1})^\T\,\MA^\T \nonumber\\
&- \MA\,\tilde{\vf}(\vx_{k-1})\,\vx_k^\T + \MA\,\tilde{\vf}(\vx_{k-1})\tilde{\vf}(\vx_{k-1})^T\,\MA^\T \mid \vy_{1:T} \big] \big\},
\end{align}
which due to linearity of expectation equals
\begin{align}
=& - \frac{T}{2} \tr \Big\{ \MQ^{-1} \Big[ \frac{1}{T}\,\sum_{k=1}^T \E[\vx_k\,\vx_k^\T \mid \vy_{1:T}]  \nonumber \\
&- \frac{1}{T}\ \big(\sum_{k=1}^T \E[\vx_k\,\tilde{\vf}(\vx_{k-1})^\T \mid \vy_{1:T}] \big) \, \MA^\T \nonumber \\
&- \MA \, \frac{1}{T}\,\sum_{k=1}^T \E[\tilde{\vf}(\vx_{k-1})\,\vx^\T \mid \vy_{1:T}]  \nonumber \\
&+ \MA \, \frac{1}{T} \, \sum_{k=1}^T \big( \E[\tilde{\vf}(\vx_{k-1})\,\tilde{\vf}(\vx_{k-1})^\T \mid \vy_{1:T}] \big) \, \MA^\T \Big] \Big\}.
\end{align}
Noting that $\E[\vx_k\,\vx_k^\T \mid \vy_{1:T}] = \vm_{k\mid T}\,\vm_{k\mid T}^\T + \MP_{k\mid T}$ and substituting in the notation introduced in Equations~(\ref{eq:MSigma}--\ref{eq:MD}), we obtain
\begin{align}
=& - \frac{T}{2} \tr \Big\{ \MQ^{-1} \Big[\MSigma - \MC\,\MA^\T - \MA\,\MC^\T + \MA\,\MPhi \, \MA^\T \Big] \Big\}.
\label{eq:appendixterm1}
\end{align}
Similar calculation for the last term in Equation~\eqref{eq:appendixQ}, noting that $\E[\vy_k\,\vy_k^\T \mid \vy_{1:T}] = \vy_k\,\vy_k^\T$, gives 
\begin{equation}
-\frac{T}{2} \tr \Big\{ \MR^{-1} \Big[ \MD - \MB\,\MH^\T - \MH\,\MB^\T + \MH \, \MTheta \, \MH\T \Big] \Big\}.
\label{eq:appendixterm2}
\end{equation}
Substituting Equations~\eqref{eq:appendixterm1} and \eqref{eq:appendixterm2} into Equation~\eqref{eq:appendixQ}, we get
\begin{align}
  &\mathcal{Q}(\vtheta,\vtheta^{(n)}) = \nonumber \\
  & - \frac{1}{2} \log |2\pi\,\MP_0|
    - \frac{T}{2} \log |2\pi\,\MQ| 
    - \frac{T}{2} \log |2\pi\,\MR| \nonumber \\
  & - \frac{1}{2} \tr \Big\{ \MP_0^{-1} \Big[
      \MP_{0 \mid T} + (\vm_{0 \mid T} - \vm_0) \, (\vm_{0 \mid T} - \vm_0)^\T
      \Big] \Big\} \nonumber \\
  & - \frac{T}{2}
     \tr \Big\{ \MQ^{-1}  \Big[ \MSigma
     -  \MC \, \MA^\T 
     - \MA \, \MC^\T 
     + \MA \, \MPhi
      \, \MA^\T
      \Big] \Big\} \nonumber \\
  &  -\frac{T}{2} 
     \tr \Big\{ \MR^{-1} \nonumber \Big[
      \MD
    -  \MB \, \MH^\T
    - \MH \, \MB^\T
    + \MH \, \MTheta \, \MH^\T
      \Big] \Big\}.
\end{align}

To maximize this with respect to the parameters $(\vm_0,\MP_0,\MA,\MH,\MQ,\MR)$ we differentiate with respect to parameter in question and set the derivative to $0$. For $\MQ$:
\begin{align}
\frac{\dd\mathcal{Q}}{\dd\MQ} =& -\frac{T}{2}\,\frac{\dd}{\dd\MQ}\log |2\pi\MQ|\nonumber \\ & - \frac{T}{2}\,\frac{\dd}{\dd\MQ  }\tr \Big\{ \MQ^{-1}  \Big[ \MSigma
     -  \MC \, \MA^\T 
     - \MA \, \MC^\T 
     + \MA \, \MPhi
      \, \MA^\T
      \Big] \Big\} \nonumber \\    
\quad =& -\frac{T}{2}\,\MQ^{-1} \nonumber \\ &+ \frac{T}{2}\,\MQ^{-1}\,\Big[ \MSigma
     -  \MC \, \MA^\T 
     - \MA \, \MC^\T 
     + \MA \, \MPhi
      \, \MA^\T
      \Big] \, \MQ^{-1}.
\end{align}
Setting the derivative equal to $\mathbf{0}$, we obtain the equation
\begin{align}
\frac{T}{2}\,\MQ^{-1} = \frac{T}{2}\,\MQ^{-1}\,\Big[ \MSigma
     -  \MC \, \MA^\T 
     - \MA \, \MC^\T 
     + \MA \, \MPhi
      \, \MA^\T
      \Big] \, \MQ^{-1}.
\end{align}
Multiplying from right by $\frac{2}{T}\,\MQ$ and from left by $\MQ$ gives
\begin{equation}
\MQ =  \MSigma
     -  \MC \, \MA^\T 
     - \MA \, \MC^\T 
     + \MA \, \MPhi
      \, \MA^\T
      .
\end{equation}
The derivations for the optimal solutions of $\MR$ and $\MP_0$ are similar. For $\MA$:
\begin{align}
\frac{\dd\mathcal{Q}}{\dd\MA} =& -\frac{T}{2}\,\big[-\frac{\dd}{\dd\MA}\tr(\MQ^{-1}\,\MC\,\MA^\T)  \nonumber \\&- \frac{\dd}{\dd\MA}\tr(\MQ^{-1}\,\MA\,\MC^T) + \frac{\dd}{\dd\MA}\tr(\MA\,\MPhi\,\MA^\T) \big] \nonumber \\
=& -\frac{T}{2}\,\MQ^{-1}\,[2\,\MA\,\MPhi - 2\MC]
\end{align}
Since $\MQ^{-1}$ is nonsingular, the derivative is zero only if the last factor is zero. If $\MPhi$ is invertible, this in turn implies
\begin{equation}
\MA = \MC\,\MPhi^{-1}.
\end{equation}
The derivations for the optimal solutions of $\MH$ and $\vm_0$ are similar. 

If the parameter $\vtheta$ is any subset of $\{\MA,\MH,\MQ,\MR,\vm_0,\MP_0\}$, it can be optimized by these closed-form expressions. First, note that $(\MA,\MQ)$, $(\MH,\MR)$ and $(\vm_0,\MP_0)$ are independent in the sense that, for example, the optimal $\MA$ and $\MQ$ do not depend on the other four parameters. Furthermore, the optimal $\MA$ does not depend on $\MQ$. Thus, $\MA$ and $\MQ$ can be jointly optimized by first solving the optimal $\MA$ and then substituting that into the expression of optimal $\MQ$. Similar reasoning works for $(\MH, \MR)$ and $(\vm_0, \MP_0)$.

\section*{Acknowledgments}
This work was supported by grants from the Academy of Finland (266940,
273475) and by the Emil Aaltonen foundation. We acknowledge the computational resources provided by the Aalto Science-IT project.

{\small
\bibliographystyle{IEEEtran}
\bibliography{IEEEfull,bibliography}
}

\end{document}